\documentclass[%
aip,
superscriptaddress,
 amsmath,amssymb,
reprint,%
]{revtex4-2}
\usepackage{amsmath, amssymb, amscd, amsthm, amsfonts}
\usepackage{graphicx}
\usepackage{dcolumn}
\usepackage{bm}
\usepackage{lineno}

\usepackage[utf8]{inputenc}
\usepackage[T1]{fontenc}
\usepackage{mathptmx}
\usepackage{bm}
\usepackage{etoolbox}
\usepackage{tabularx}
\usepackage{subcaption}
\usepackage{svg}
\usepackage{epstopdf}
\usepackage[justification=raggedright]{caption}
\usepackage{lmodern}
\usepackage{multirow}
\usepackage{textcomp,mathcomp}
\usepackage{booktabs}  
\usepackage{cellspace} 
\usepackage{array}      
\usepackage{makecell}
\usepackage{cancel}
\usepackage{CJKutf8}

\usepackage{hyperref}
\hypersetup{
colorlinks=true,
linkcolor=blue,
anchorcolor=blue,
citecolor=blue}

\makeatletter
\def\@email#1#2{%
 \endgroup
 \patchcmd{\titlejamming@produce}
  {\frontmatter@RRAPformat}
  {\frontmatter@RRAPformat{\produce@RRAP{*#1\href{mailto:#2}{#2}}}\frontmatter@RRAPformat}
  {}{}
}%
\makeatother

\preprint{AIP/123-QED}

\begin{document}

\title{Finite-inertia effects in Langevin dynamics of a lopsided elastic dumbbell using exponential-time differencing schemes}

\author{Lei Song}
\affiliation{State Key Laboratory of Fluid Power and Mechatronic Systems and Department of Engineering Mechanics, Zhejiang University, Hangzhou 310027, China}

\author{Dingyi Pan}
\thanks{Corresponding author: dpan@zju.edu.cn}
\affiliation{State Key Laboratory of Fluid Power and Mechatronic Systems and Department of Engineering Mechanics, Zhejiang University, Hangzhou 310027, China}
\affiliation{Zhejiang Key Laboratory of Intelligent High Speed UAV Technology, Hangzhou 310027, China}
\affiliation{Innovation Center of Yangtze River Delta, Zhejiang University, Jiaxing 314102, China}

\author{Nhan Phan-Thien}
\affiliation{State Key Laboratory of Fluid Power and Mechatronic Systems and Department of Engineering Mechanics, Zhejiang University, Hangzhou 310027, China}
\affiliation{Department of Mechanical Engineering, National University of Singapore, Singapore 117575}
\affiliation{Mekong University (MKU), Vinh Long Province, Vietnam}	

\date{\today}

\begin{abstract}
Inertia effects in the Langevin dynamics of a lopsided elastic dumbbell are investigated using exponential-time-differencing (ETD) integrators for the corresponding stiff stochastic equations at small mass limit. Starting from the bead-level underdamped Langevin model, we formulate the dynamics in modal coordinates, highlighting two distinct friction scales: an additive friction $\zeta_{\rm trans}=\zeta_1+\zeta_2$ controlling translation ($\zeta_i, i=1,2$ are the friction factor on bead $i$), and an effective internal friction $1/\zeta_{\rm eff}=1/\zeta_1+1/\zeta_2$ controlling configurational relaxation, with relaxation time $\tau_R=\zeta_{\rm eff}/H$ for a Hookean spring of stiffness $H$. We benchmark ETD against Euler--Maruyama and overdamped Brownian dynamics using equilibrium statistics, time-domain autocorrelations, and frequency-domain power spectra of the end-to-end vector. When time is rescaled by $\tau_R$, configurational and orientational relaxation curves collapse across asymmetry ratios, showing that the dominant long-time structural dynamics remains close to the overdamped description. Inertial signatures are instead confined to short-time transients, high-frequency modifications of the configurational spectrum, and a transient coupling between translational and internal modes. This study provides a practical and accurate route for lopsided dumbbells across overdamped and weakly underdamped regimes, and clarify how mass and friction asymmetry affect the translational and internal dynamics.
\end{abstract}

\maketitle

\section{Introduction}\label{section-introduction}
Langevin dynamics provides a standard microscopic description for the stochastic motion of suspended particles in a viscous solvent. \cite{langevin1908theorie,uhlenbeck1930theory,Dingyi2025} In polymer and soft-matter physics, bead-spring (including dumbbell) models are commonly formulated within this framework and serve as minimal descriptions of configurational fluctuations, orientational relaxation, and rheological response. \cite{doi1988theory,bird1987dynamics,serafini2024polymers} In much of the classical Hookean dumbbell theory, however, the analysis is carried out in the overdamped limit of the Langevin equation. In this case, bead inertia is neglected and momentum variables are eliminated.\cite{doi1988theory,bird1987dynamics,Dingyi2025} This reduction underlies the standard Smoluchowski description and leads to the familiar Ornstein--Uhlenbeck dynamics for the relative coordinate. \cite{uhlenbeck1930theory,doi1988theory,bird1987dynamics}

The overdamped approximation is often highly successful because momentum typically relaxes much faster than the configurational degrees of freedom, so that the approximation captures the dominant long-time structural relaxation and enables many of the standard analytical results used in polymer dynamics. \cite{doi1988theory,bird1987dynamics,zwanzig2001nonequilibrium,pathria2017statistical} At the same time, removing the mass terms also removes the fast inertial stage of the dynamics. Consequently, although finite inertia is small, it may still leave measurable signatures in short-time correlations, high-frequency spectra, and phase-space cross-correlations. \cite{Dingyi2025,gao2024some} Indeed, inertia effects in harmonic dumbbell models {, i.e., dumbbells connected by a Hookean (linear) spring with quadratic potential,} have been analyzed in the context of linear viscoelasticity, where the momentum and bond-relaxation modes were shown to be kinetically coupled and weak inertia was found to modify bond relaxation when momentum relaxation is not sufficiently fast. \cite{uneyama2019inertia} This work examines, within a Langevin framework, how such finite-inertia effects appear.%

This effect arises in a \emph{lopsided} dumbbell, in which the two beads have unequal sizes and therefore unequal masses and Stokes drag coefficients. Recent work has extended dumbbell descriptions to lopsided rigid and elastic particles, clarifying how asymmetry modifies transport and rheological behavior. \cite{phan2024lopsidedtutorial,phan2024lopsided} These \emph{lopsided} dumbbell ideas have also begun to appear in more application-oriented rheological settings.~\cite{kanso2025rheology} In a lopsided dumbbell, translational and internal modes need not respond on the same inertial timescale, so the effect of neglecting the mass terms need not be uniform across observables. In these recent developments, a systematic study of finite-inertia effects for \emph{lopsided} elastic dumbbells still remains limited. In particular, while inertia effects have been analyzed for harmonic dumbbell models, especially in the context of linear viscoelasticity, \cite{uneyama2019inertia} a quantitative characterization of how mass and drag asymmetry enter short-time correlations, phase-space cross-correlations, and frequency-domain responses is still lacking.

When the bead masses are finitely small, the underdamped Langevin equations contain widely separated momentum and configurational timescales and become stiff, in the sense that explicit time-stepping is constrained by the fastest relaxation scale even when the main observables evolve much more slowly. \cite{kloeden1977numerical,kloeden1992introduction,Dingyi2025} This regime is relevant not only when the particle mass cannot be entirely ignored, but also when one wishes to quantify the effect of the small mass omitted in the overdamped theory. 
While several effective integrators exist for underdamped stochastic dynamics, our adoption of exponential-time-differencing (ETD) methods is motivated by the fact that closely related ETD ideas have already been successfully applied to low-mass mesoscopic particle systems, in particular dissipative particle dynamics (DPD)\cite{phan2014exponential,mai2013dissipative}{, which also leads to stiff stochastic differential equations in the small-mass limit with a similar mathematical structure.}
More broadly, ETD methods were developed for stiff deterministic systems and continue to be refined for stiff problems with nontrivial linear parts. \cite{cox2002exponential,permyakova2025high}

In this work, we study finite-inertia effects in the Langevin dynamics of a lopsided Hookean dumbbell, employ ETD integrators for the stiff underdamped equations, and benchmark the results against the Euler--Maruyama (EM) method and overdamped Brownian dynamics (BD) in both the time and frequency domains.

The remainder of the paper is organized as follows. Section~\ref{sec:method} introduces the underdamped dumbbell model and the mode decomposition used throughout. The ETD integrators and the sampling of the associated stochastic increments are then presented. Section~\ref{sec:results} reports numerical results, including equilibrium statistics, correlation functions, and spectral characterizations of configurational fluctuations, and quantifies the regimes where finite inertia is detectable and where ETD offers practical advantages over EM. Section~\ref{sec:conclusion} summarizes the main findings and discusses extensions.



\section{Langevin theory for a lopsided dumbbell and ETD integration}\label{sec:method}
\subsection{Langevin description of a lopsided dumbbell}
The lopsided dumbbell framework follows the recent rigid and elastic formulations introduced in Refs.,~\cite{phan2024lopsidedtutorial,phan2024lopsided} but here we retain finite bead inertia and analyze short-time and frequency-domain deviations from the overdamped description.
First, we consider a lopsided dumbbell as shown in Fig.~\ref{fig.lopsideddumbbell}, where the forces acting on the beads can be written as:
\begin{figure}[tbh]\centering
	\includegraphics[width=0.85\linewidth]{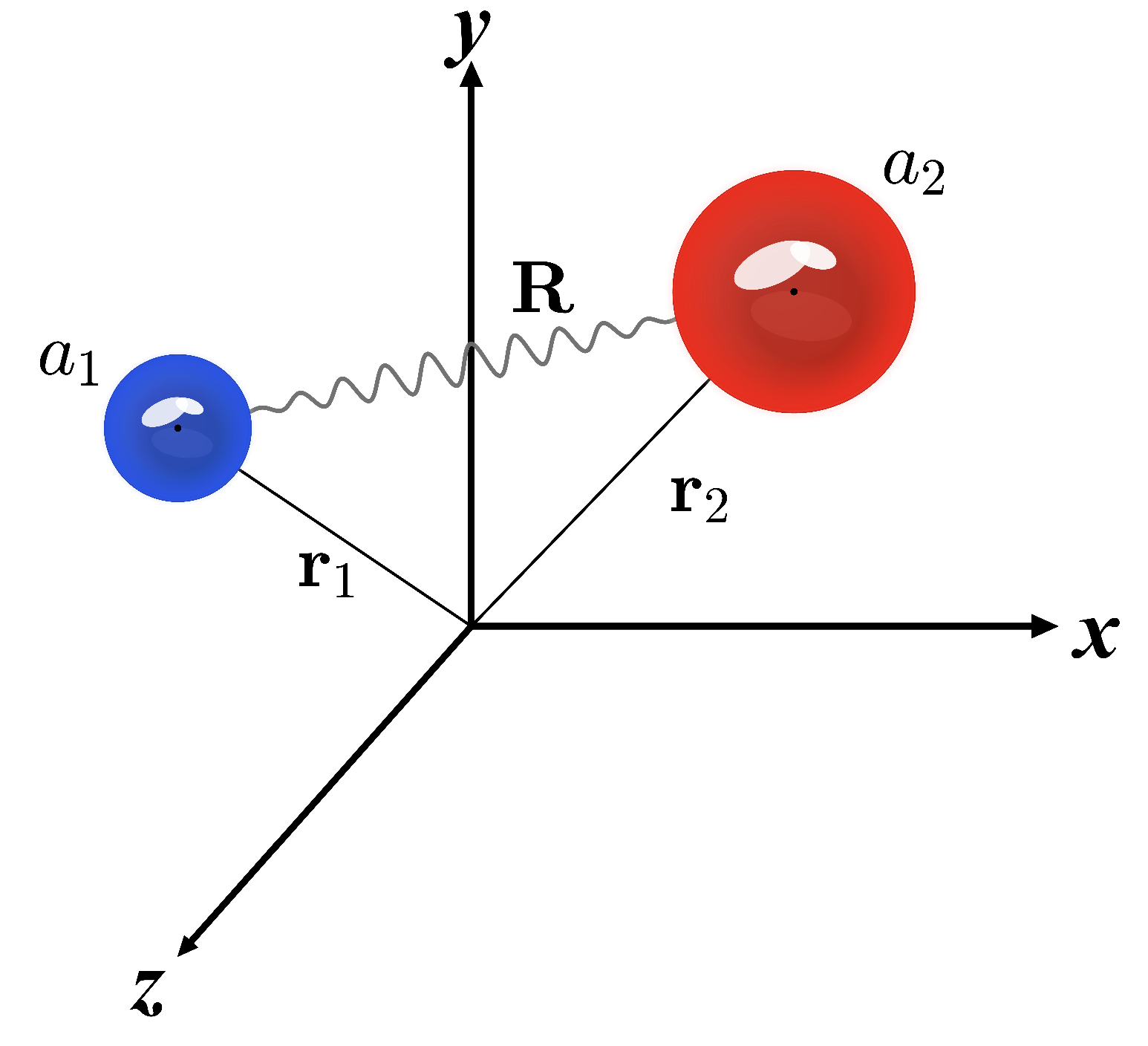}
    \caption{A lopsided elastic dumbbell~\cite{phan2024lopsided,Dingyi2025} consists of two unequal-sized beads of radii $a_1$ and $a_2$, whose positions are denoted by $\mathbf r_1$ and $\mathbf r_2$. The beads are connected by a spring, and the end-to-end connector vector is $\mathbf R=\mathbf r_2-\mathbf r_1$.}
	\label{fig.lopsideddumbbell}
\end{figure}
\begin{equation}
	\begin{aligned}
		m_1\ddot{\mathbf r_1}+\zeta_1\dot{\mathbf r_1}=\zeta_1\mathbf u_1+H(\mathbf r_2-\mathbf r_1)+\mathbf F_1^{(b)}(t),
		\\
		m_2\ddot{\mathbf r_2}+\zeta_2\dot{\mathbf r_2}=\zeta_2\mathbf u_2+H(\mathbf r_1-\mathbf r_2)+\mathbf F_2^{(b)}(t),
	\end{aligned}
    \label{eq:langevin}
\end{equation}
where $\mathbf u_i=\mathbf u(\mathbf r_i)$ is the imposed fluid velocity evaluated at bead $i$, and ${\bf F}^{(b)}_i$ is the Brownian force on bead $i$ satisfying the fluctuation--dissipation theorem (see, for example, \cite{zwanzig2001nonequilibrium,phan2024lopsidedtutorial,Dingyi2025}):
\begin{equation}
\begin{aligned}
	\langle {\mathbf F}^{(b)}_i(t)\rangle &= 0,
    \\
    \langle {\mathbf F}^{(b)}_i (t) {\bf F}^{(b)}_j(t+\tau) \rangle &= 2k_BT\zeta_i\delta_{ij}\delta(\tau) {\bf I}.
	\label{Eq18}
\end{aligned}
\end{equation}
{In the present free-draining model, hydrodynamic interactions between beads are neglected~\cite{adler1980polymer,stepto2015definitions}, so each bead experiences only local Stokes drag and a Brownian force with constant amplitude.
Consequently, the diffusion tensor is state-independent and is chosen according to Eq.~\eqref{Eq18} to satisfy the fluctuation--dissipation theorem. 
Thus, the It\^o and Stratonovich interpretations give the same stochastic dynamics here~\cite{ito1944109,ito1951stochastic,kloeden1992introduction}; this point is mentioned only for completeness.}

We next introduce the center of resistance $\mathbf Q$ (which, under the present free-draining approximation, coincides with the center of friction/drag) satisfying $(\zeta_1+\zeta_2)\mathbf Q=\zeta_1\mathbf r_1+\zeta_2\mathbf r_2$, and the dumbbell vector $\mathbf R=\mathbf r_2-\mathbf r_1$, and Taylor series expand the fluid velocity about the center of resistance,
\begin{equation}
	\begin{aligned}
	&\mathbf u_i=\mathbf u^{(c)}+(\mathbf {r}_i-\mathbf Q)\cdot \nabla\mathbf u^{(c)}
	\\
	&+\frac{1}{2}(\mathbf {r}_i-\mathbf Q)(\mathbf {r}_i	-\mathbf Q):\nabla\nabla\mathbf u^{(c)}+\cdots.
	\end{aligned}
\end{equation}
Since $\mathbf{r}_1-\mathbf Q=-\mathbf R^{(1)},\quad \mathbf{r}_2-\mathbf Q=\mathbf R^{(2)}$, and ${\bf R}^{(1)} + {\bf R}^{(2)} = {\bf R}$, the Langevin equations become, ignoring higher order terms,
\begin{equation}
\begin{aligned}
M_{11}\ddot{\mathbf Q} +& M_{12}\ddot{\mathbf R} + \dot{\mathbf Q}= \\
&\mathbf u^{(c)} + (\zeta_1+\zeta_2)^{-1}\mathbf F^{(b,cr)}(t) ,
\\
M_{21}\ddot{\mathbf Q} +& M_{22}\ddot{\mathbf R} + \dot{\mathbf R}= \\
&\mathbf L\cdot \mathbf R - 
\frac{\zeta_1+\zeta_2}{\zeta_1\zeta_2}H\mathbf R + \frac{1}{\zeta_2}\mathbf F^{(b,R)}(t),
\end{aligned}
\label{eq.dumbbelllangevin}
\end{equation}
with
\begin{equation}
	\begin{aligned}
M_{11}&=\frac{m_1+m_2}{\zeta_1+\zeta_2},
\\
M_{12}&=\frac{m_2\zeta_1-m_1\zeta_2}{(\zeta_1+\zeta_2)^2},
\label{eq:M11M12}
	\end{aligned}
\end{equation}
and
\begin{equation}
\begin{aligned}
M_{21}&=\frac{m_2}{\zeta_2}-\frac{m_1}{\zeta_1},
\\
M_{22}&=\frac{\zeta_1 m_2}{\zeta_2(\zeta_1+\zeta_2)}
+\frac{\zeta_2 m_1}{\zeta_1(\zeta_1+\zeta_2)}.
\end{aligned}
\label{eq:M21M22}
\end{equation}
Here, $\mathbf F^{(b,cr)}(t)=\mathbf F_1^{(b)}(t)+\mathbf F_2^{(b)}(t)$ is the Brownian force exerted on the center of resistance, $\mathbf F^{(b,R)}(t)=\mathbf F_2^{(b)}(t)-\frac{\zeta_2}{\zeta_1}\mathbf F^{(b)}_1(t)$ is the Brownian force acting on the center-to-center vector, $\mathbf L=\nabla \mathbf{u^{(c)}}^T$ is the velocity gradient evaluated at the center of resistance. The mass tensor has a positive determinant, as required for a physical system with positive kinetic energy. Indeed, after some algebra,
\begin{align}
    \det \left( {\bf M} \right) &= M_{11}M_{22} - M_{12}M_{21} \nonumber \\
    &=\frac{1}{(\zeta_1 + \zeta_2)^2}
\left[
m_1 m_2
\left(
\frac{\zeta_1}{\zeta_2} + \frac{\zeta_2}{\zeta_1} + 2
\right)
\right].
\end{align}
Then, using the identity
\[
\frac{\zeta_1}{\zeta_2} + \frac{\zeta_2}{\zeta_1} + 2
= \frac{(\zeta_1 + \zeta_2)^2}{\zeta_1 \zeta_2},
\]
we obtain 
\begin{equation}   
\det \left( {\bf M} \right) = \frac{m_1 m_2}{\zeta_1 \zeta_2},
\end{equation}
which is recognized as the product of two inertial time scales $\tau_{m,i} = \frac{m_i}{\zeta_i}, i=1,2$ (no sum on $i$). 
Since $\det \left( {\bf M} \right)>0$ for $m_i,\zeta_i>0$, the inertial matrix is nonsingular for all physically admissible parameters. This guarantees the well-posedness of the short-time dynamics and excludes any degeneracy of the inertial modes.
In addition, the off-diagonal terms, $M_{12}$ and $M_{21}$, represent the coupling between the translational and internal modes; for a symmetric dumbbell, these terms vanish identically and thus the symmetric dumbbell exactly decouples in these normal mode variables.  

\subsection{ETD integration for the inertial bead dynamics}

To retain small but nonzero mass effects omitted in the overdamped theory, we work with the bead-level underdamped Langevin equations. These equations contain the linear viscous term $-(\zeta_i/m_i)\mathbf v_i$ (no sum), which becomes stiff for small bead mass.\cite{kloeden1977numerical,kloeden1992introduction,phan2014exponential}
To resolve this fast relaxation, we apply exponential time differencing (ETD), following the general ETD idea for stiff systems and its low-mass stochastic adaptation in mesoscopic particle dynamics, \cite{cox2002exponential,phan2014exponential,permyakova2025high} to the velocity equation of bead $i$,
\begin{equation}
\begin{aligned}
\frac{d\mathbf r_i}{dt}&=\mathbf v_i,
\\
\frac{d\mathbf v_i}{dt}+c_{i}\mathbf v_i
&=\frac{1}{m_i}\mathbf F_i^{(c)}(t)+\frac{1}{m_i}\mathbf F_i^{(b)}(t),
\end{aligned}
\label{eq:vi_langevin_ci}
\end{equation}
where $c_{i}=\zeta_i/m_i = 1/\tau_{(m,i)}$ (no sum on $i$) is a {\it scalar}. The variation-of-constants formula over one time step gives (no sum on $i$)
\begin{equation}
\begin{aligned}
&\mathbf v_i^{\,n+1} =e^{-c_{i}h}\mathbf v_i^{\,n}
\\
&+\frac{1}{m_i}\int_0^{h}e^{-c_{i}(h-s)}\mathbf F_i^{(c)}(t_n+s)\,ds+\mathbf G_i^n, 
\end{aligned}
\label{eq:etd_v_general_short}
\end{equation}
where $h$ is time step and $\mathbf G_i^n$ denotes the Brownian contribution integrated over the step.
The position is updated according to
\begin{equation}
\mathbf r_i^{\,n+1}
=\mathbf r_i^{\,n}
+\int_0^{h}\mathbf v_i(t_n+s)\,ds.
\label{eq:etd_r_general_short}
\end{equation}
For compactness, we define (no sum on $i$)
\begin{equation}
\begin{aligned}
\psi_{1,i}&=\frac{1-e^{-c_ih}}{c_i},\\
\psi_{2,i}&=\frac{c_ih-1+e^{-c_ih}}{c_i^2},\\
\psi_{3,i}&=\frac{(c_ih)^2/2-c_ih+1-e^{-c_ih}}{c_i^3}.
\end{aligned}
\end{equation}

We refer to ETD1 when the deterministic force is held constant over a time step, $\mathbf F_i^{(c)}(t_n+s)\approx \mathbf F_i^{\mathrm c,n}$, yielding (no sum on $i$)
\begin{align}
\mathbf v_i^{\,n+1}
&=e^{-c_ih}\mathbf v_i^{\,n}
+\frac{\psi_{1,i}}{m_i}\mathbf F_i^{\mathrm c,n}
+\mathbf G_i^n,
\label{eq:ETD1vel_short}\\
\mathbf r_i^{\,n+1}
&=\mathbf r_i^{\,n}
+\psi_{1,i}\mathbf v_i^{\,n}
+\frac{\psi_{2,i}}{m_i}\mathbf F_i^{\mathrm c,n}
+\mathbf H_i^n,
\label{eq:ETD1pos_short}
\end{align}
where $\mathbf H_i^n$ is the corresponding Brownian increment in the position update.

Meanwhile, we refer to ETD2 when the deterministic force is approximated linearly over the step,
\begin{equation}
\mathbf F_i^{(c)}(t_n+s)
\approx
\mathbf F_i^{\mathrm c,n}
+\frac{s}{h}
\left(\mathbf F_i^{\mathrm c,*}-\mathbf F_i^{\mathrm c,n}\right),
\end{equation}
where $\mathbf F_i^{\mathrm c,*}$ is evaluated from a predictor state. This gives (no sum on $i$)
\begin{equation}
\begin{aligned}
\mathbf v_i^{\,n+1}
&=e^{-c_ih}\mathbf v_i^{\,n}
\\
&+\frac{\psi_{1,i}}{m_i}\mathbf F_i^{\mathrm c,n}
+\frac{\psi_{2,i}}{m_ih}
\left(\mathbf F_i^{\mathrm c,*}-\mathbf F_i^{\mathrm c,n}\right)
+\mathbf G_i^n,
\end{aligned}\label{eq:ETD2vel_short}
\end{equation}
\begin{equation}
\begin{aligned}
    \mathbf r_i^{\,n+1}
&=\mathbf r_i^{\,n}+\psi_{1,i}\mathbf v_i^{\,n}
\\
&+\frac{\psi_{2,i}}{m_i}\mathbf F_i^{\mathrm c,n}
+\frac{\psi_{3,i}}{m_ih}
\left(\mathbf F_i^{\mathrm c,*}-\mathbf F_i^{\mathrm c,n}\right)
+\mathbf H_i^n.
\end{aligned}\label{eq:ETD2pos_short}
\end{equation}
Therefore, ETD1 and ETD2 correspond to first- and second-order approximations of the deterministic force variation over a single timestep. Accordingly, for sufficiently smooth deterministic forcing, they are expected to attain weak orders 1 and 2, respectively.

The zero-mean Brownian increments $\mathbf G_i^n$ and $\mathbf H_i^n$ are
jointly Gaussian, and correlated component-wise. For each Cartesian component,
\begin{equation}
\mathrm{Cov}
\begin{pmatrix}
G\\
H
\end{pmatrix}
=
\frac{k_B T}{m_i}
\begin{pmatrix}
\sigma_{vv}^{(i)} & \sigma_{vr}^{(i)}\\
\sigma_{vr}^{(i)} & \sigma_{rr}^{(i)}
\end{pmatrix},
\label{eq:cov_GR_compact}
\end{equation}
with (no sum on $i$)
\begin{equation}
\begin{aligned}
\sigma_{vv}^{(i)}&=1-e^{-2c_ih},\\
\sigma_{vr}^{(i)}&=\frac{(1-e^{-c_ih})^2}{c_i},\\
\sigma_{rr}^{(i)}&=\frac{2c_ih-3+4e^{-c_ih}-e^{-2c_ih}}{c_i^2}.
\end{aligned}
\end{equation}
These increments are sampled numerically from the covariance, 
Eq.~\eqref{eq:cov_GR_compact}.

To assess the ETD integrator, we ran simulations with EM, ETD1, and ETD2 under the same conditions. 
{The EM scheme used here corresponds to a velocity-Verlet-type discretization of the underdamped Langevin equation, rather than a simple explicit Euler–Maruyama method.}
We compared three ensemble quantities: $k_BT$, $\langle R^2\rangle$, and the pressure. The comparisons of the three simulation parameters are listed in three tables below (Tables~\ref{tab:tempVerify}--\ref{tab:pressVerify}).
\begin{table}[htbp]
\centering
\caption{Equilibrium verification of the kinetic temperature for EM, ETD1, and ETD2 at different timesteps $h$. The theoretical value is $k_B T=1$.}
\label{tab:tempVerify}
\renewcommand{\arraystretch}{1.3}           
\begin{ruledtabular}
\begin{tabular}{ccccccc}
\multirow{2}{*}{$h$} & \multicolumn{2}{c}{EM}                                                         & \multicolumn{2}{c}{ETD1}                                                       & \multicolumn{2}{c}{ETD2}                                                       \\ \cline{2-7} 
                            & $\langle k_BT\rangle$ & \begin{tabular}[c]{@{}c@{}}Error\\ ($\%$)\end{tabular} & $\langle k_BT\rangle$ & \begin{tabular}[c]{@{}c@{}}Error\\ ($\%$)\end{tabular} & $\langle k_BT\rangle$ & \begin{tabular}[c]{@{}c@{}}Error\\ ($\%$)\end{tabular} \\ \hline
0.0002                      & 1.0019                & 0.19                                                   & 0.9988                & 0.12                                                   & 0.9987                & 0.13                                                   \\
0.0004                      & 1.0013                & 0.13                                                   & 0.9983                & 0.17                                                   & 0.9981                & 0.19                                                   \\
0.0008                      & 1.0018                & 0.18                                                   & 0.9986                & 0.14                                                   & 0.9984                & 0.16                                                   \\
0.002                       & 0.9983                & 0.17                                                   & 0.9986                & 0.14                                                   & 0.9983                & 0.17                                                   \\
0.004                       & 0.9999                & 0.01                                                   & 0.9981                & 0.19                                                   & 0.9981                & 0.19                                                   \\
0.008                       & -                     & -                                                      & 0.9982                & 0.18                                                   & 0.9982                & 0.18                                                   \\
0.02                        & -                     & -                                                      & 0.9983                & 0.17                                                   & 0.9982                & 0.18                                                   \\
0.04                        & -                     & -                                                      & 0.9982                & 0.18                                                   & 0.9983                & 0.17                                                  
\end{tabular}
\end{ruledtabular}
\end{table}
\begin{table}[htbp]
\centering
\caption{Equilibrium verification of the mean-squared connector length for EM, ETD1, and ETD2 at different timesteps $h$. The theoretical value is $\langle R^2\rangle=3$.}
\label{tab:R2Verify}
\renewcommand{\arraystretch}{1.3}           
\begin{ruledtabular}
\begin{tabular}{ccccccc}
\multirow{2}{*}{$h$} & \multicolumn{2}{c}{EM}                                                        & \multicolumn{2}{c}{ETD1}                                                      & \multicolumn{2}{c}{ETD2}                                                      \\ \cline{2-7} 
                            & $\langle R^2\rangle$ & \begin{tabular}[c]{@{}c@{}}Error\\ ($\%$)\end{tabular} & $\langle R^2\rangle$ & \begin{tabular}[c]{@{}c@{}}Error\\ ($\%$)\end{tabular} & $\langle R^2\rangle$ & \begin{tabular}[c]{@{}c@{}}Error\\ ($\%$)\end{tabular} \\ \hline
0.0002                      & 3.0148               & 0.49                                                   & 2.9839               & 0.54                                                   & 3.0032               & 0.11                                                   \\
0.0004                      & 3.0022               & 0.07                                                   & 3.0076               & 0.25                                                   & 3.0008               & 0.03                                                   \\
0.0008                      & 2.9907               & 0.31                                                   & 2.9872               & 0.43                                                   & 3.0033               & 0.11                                                   \\
0.002                       & 3.0043               & 0.14                                                   & 3.0070               & 0.23                                                   & 3.0031               & 0.10                                                   \\
0.004                       & 3.9342               & 31.14                                                  & 3.0095               & 0.32                                                   & 3.0041               & 0.14                                                   \\
0.008                       & -                    & -                                                      & 3.0190               & 0.63                                                   & 2.9991               & 0.03                                                   \\
0.02                        & -                    & -                                                      & 3.0362               & 1.21                                                   & 3.0019               & 0.06                                                   \\
0.04                        & -                    & -                                                      & 3.0789               & 2.63                                                   & 3.0075               & 0.25                                                  
\end{tabular}
\end{ruledtabular}
\end{table}
\begin{table}[htbp]
\centering
\caption{Equilibrium verification of the pressure for EM, ETD1, and ETD2 at different timesteps $h$. The theoretical value is $p=n_d k_B T$.}
\label{tab:pressVerify}
\renewcommand{\arraystretch}{1.3}           
\begin{ruledtabular}
\begin{tabular}{ccccccc}
\multirow{2}{*}{$h$} & \multicolumn{2}{c}{EM}                                                      & \multicolumn{2}{c}{ETD1}                                                    & \multicolumn{2}{c}{ETD2}                                                    \\ \cline{2-7} 
                            & $\langle p\rangle$ & \begin{tabular}[c]{@{}c@{}}Error\\ ($\%$)\end{tabular} & $\langle p\rangle$ & \begin{tabular}[c]{@{}c@{}}Error\\ ($\%$)\end{tabular} & $\langle p\rangle$ & \begin{tabular}[c]{@{}c@{}}Error\\ ($\%$)\end{tabular} \\ \hline
0.0002                      & 1.1919             & 0.15                                                   & 1.1968             & 0.26                                                   & 1.1889             & 0.40                                                   \\
0.0004                      & 1.1957             & 0.17                                                   & 1.1864             & 0.61                                                   & 1.1886             & 0.42                                                   \\
0.0008                      & 1.2015             & 0.66                                                   & 1.1950             & 0.11                                                   & 1.1882             & 0.46                                                   \\
0.002                       & 1.1876             & 0.51                                                   & 1.1871             & 0.55                                                   & 1.1881             & 0.47                                                   \\
0.004                       & 0.8350             & 30.05                                                  & 1.1851             & 0.72                                                   & 1.1871             & 0.55                                                   \\
0.008                       & -                  & -                                                      & 1.1815             & 1.02                                                   & 1.1895             & 0.35                                                   \\
0.02                        & -                  & -                                                      & 1.1750             & 1.56                                                   & 1.1883             & 0.45                                                   \\
0.04                        & -                  & -                                                      & 1.1577             & 3.01                                                   & 1.1863             & 0.62                                                  
\end{tabular}
\end{ruledtabular}
\end{table}
For equilibrium validation, we compare EM, ETD1, and ETD2 for the symmetric case $\nu=1$ over a range of timesteps $h$. The simulations use the same physical parameters as in the production runs (Sec.~\ref{sec:methods}); see Table~\ref{tab:parameters} for details. 

All three schemes reproduce the target temperature within statistical uncertainty throughout their stable timestep ranges. 
{The kinetic temperature constrains only the velocity (momentum) distribution and does not probe the configurational statistics, and therefore cannot by itself guarantee correct sampling of the equilibrium ensemble. In contrast, $\langle R^2\rangle$ and $p$ involve configurational and mixed contributions, and therefore provide more discriminating tests for equilibrium accuracy.}
By contrast, $\langle R^2\rangle$ and $p$ are much more sensitive to timestep size and therefore provide more discriminating tests for equilibrium accuracy. As $h$ approaches the momentum-relaxation timescale, EM rapidly loses accuracy and breaks down at $h=0.004$, which is close to $2\tau_m$ for the present parameters, indicating that the instability is governed by this scale.
ETD1 substantially enlarges the stability window, but its errors in $\langle R^2\rangle$ and $p$ increase systematically with timestep. ETD2, in contrast, maintains consistently low errors across the entire tested range, remaining close to the theoretical equilibrium values even at large timesteps. 
{For example, at $h=0.004$ the EM scheme exhibits an error of approximately $31\%$ in $\langle R^2\rangle$, whereas ETD2 remains accurate within $0.3\%$.}
Although short-time and high-frequency observables still require a sufficiently small timestep to resolve the momentum-relaxation dynamics, ETD2 remains preferable because it reduces stiffness-induced stability restrictions and equilibrium bias in the low-mass regime. We therefore adopt ETD2 in all subsequent simulations.

\subsection{Simulation Model and Methods}
\label{sec:methods}

We simulate lopsided elastic dumbbells, we performed coarse-grained simulations using the Large-scale Atomic/Molecular Massively Parallel Simulator (LAMMPS) package.~\cite{LAMMPS} We consider a dilute solution of non-interacting dumbbells in a three-dimensional periodic simulation box. Excluded volume interactions between beads are turned off (ideal $\theta$-solvent conditions),\cite{doi1988theory} ensuring that the dynamics are governed solely by spring force, viscous drag, and thermal fluctuations from the solvent. Each dumbbell comprises two beads connected by a Hookean spring with potential energy $U(R)=\frac{1}{2}H R^2$. 
The radii of the two beads are $a_1$ and $a_2$, and the lopsidedness is characterized by the asymmetry ratio
\begin{equation}
    \nu = \frac{a_2}{a_1}.\label{eq:nu}
\end{equation}
Assuming a uniform mass density $\rho_p$, the bead masses are $m_i = \frac{4}{3}\pi\rho_p a_i^3$. The solvent is modeled as a thermal bath with temperature $T$ and viscosity $\eta_f$. Under this free-draining approximation, the friction coefficient follows Stokes' law, $\zeta_i = 6\pi\eta_f a_i$. The basic simulation parameters are summarized in Table~\ref{tab:parameters}.
\begin{table}[tbp]
	\centering
    \caption{Fixed simulation parameters used in the simulations.}
	\label{tab:parameters}
    \renewcommand{\arraystretch}{1.3}           
    \begin{ruledtabular}
	\begin{tabular}{lcc}
		Parameter & Symbol & Value \\
		\hline
		Small bead radius & $a_1$ & 0.1 \\
		Solvent viscosity & $\eta_f$ & 1.0 \\
		Bead density & $\rho_p$ & 1.0 \\
		Spring constant & $H$ & 1.0 \\
		Temperature & $k_BT$ & 1.0 \\
		Time step & $h$ & $1\times10^{-4}\sim 5\times10^{-3}$ \\
		Nominal volume fraction & $\phi$ & 0.01 \\
	\end{tabular}
    \end{ruledtabular}
\end{table}
\begin{figure}[tbh]\centering
	\includegraphics[width=\linewidth]{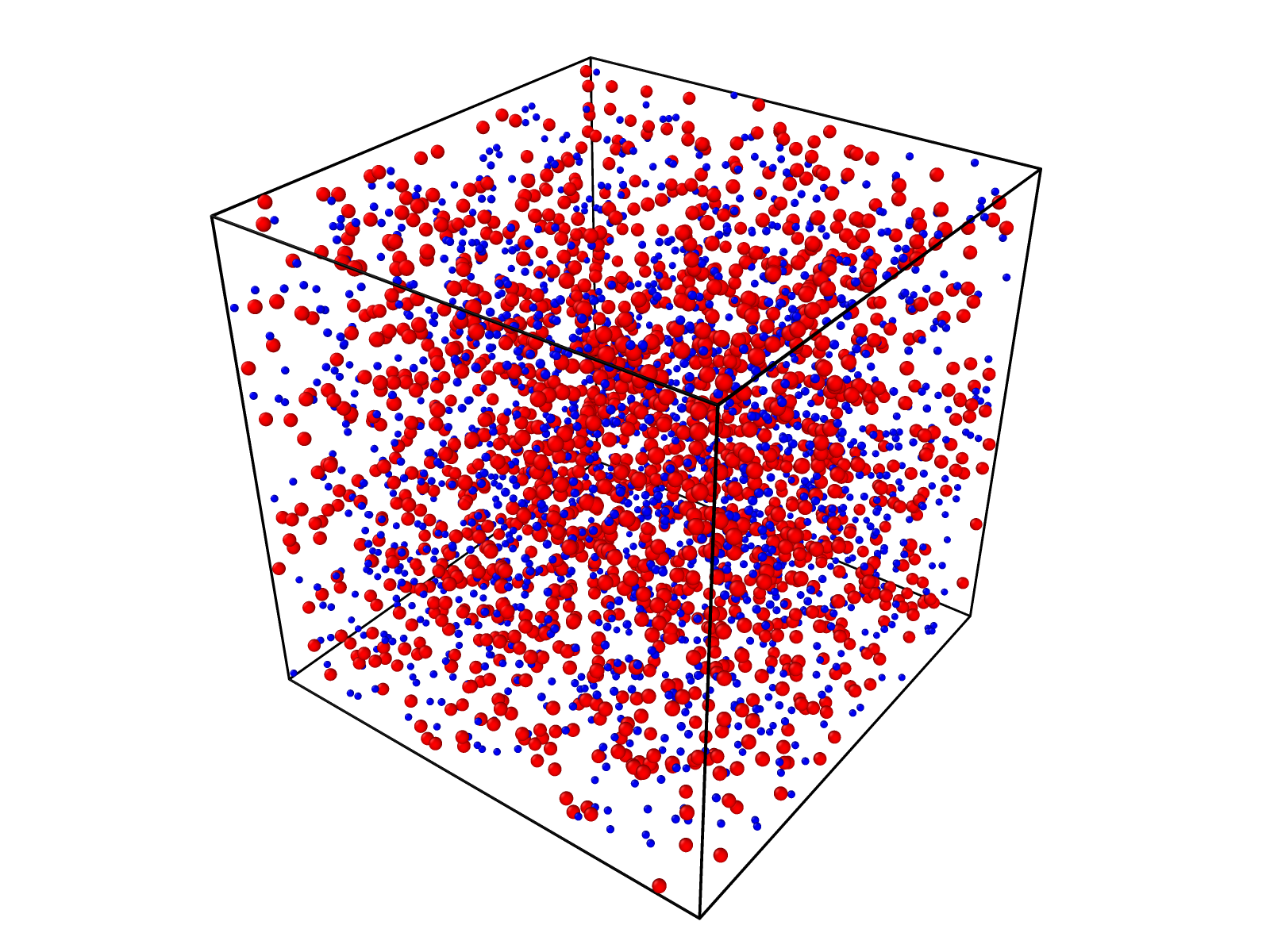}
	\caption{Schematic of a dilute solution of lopsided elastic dumbbells. The large sized bead (red) and small sized bead (blue) are connected by a virtual Hookean spring.}
	\label{fig.dumbbellsuspension}
\end{figure}

To examine both the overdamped and inertial regimes, we employ three numerical schemes: BD for the overdamped limit ($m_i \to 0$), while the Euler--Maruyama (EM) and ETD schemes are applied to the underdamped Langevin equation. In the ETD family, we use the second-order variant ETD2, which treats the linear viscous relaxation exactly and is therefore well suited to the stiff low-mass regime.
{The role of ETD here is not to remove the need to resolve the shortest inertial timescale. Instead, it provides a more robust integrator for the stiff low-mass regime. This leads to a substantially enlarged stable timestep range and reduced equilibrium bias, while still requiring sufficient resolution of the fastest inertial dynamics when short-time observables are of interest.}
In the present simulations, however, several observables probe the short-time inertial response directly, so the timestep is set with reference to the smallest momentum-relaxation time $\tau_{m,1}=m_1/\zeta_1$ (approximately $2\times 10^{-3}$ for the parameters in Table~\ref{tab:parameters}) as the relevant short-time scale. For each asymmetry ratio $\nu$ in the range $[1, 10]$, the simulation box is sized to achieve a dilute solution condition, typically containing $4000\sim8000$ dumbbells to ensure sufficient statistics.
A schematic of the model is shown in Fig.~\ref{fig.dumbbellsuspension}.

\section{Numerical Results}\label{sec:results}
\subsection{Translational dynamics}

\paragraph*{Long-time diffusivity and mean-squared displacement.}
We first examine the translational dynamics of the lopsided dumbbell. 
In the absence of an external flow, Eq.~\eqref{eq.dumbbelllangevin} reduces to linear equations for the coupled $\mathbf Q$ and $\mathbf R$ dynamics. Writing the corresponding equations for the velocities $\mathbf V_Q=\dot{\mathbf Q}$ and $\mathbf V_R=\dot{\mathbf R}$ in Fourier space, we obtain
\begin{equation}
	\begin{pmatrix}
		i\omega M_{11}+1 & i\omega M_{12} \\
		i\omega M_{21} & i\omega M_{22}+1+\dfrac{\kappa}{i\omega}
	\end{pmatrix}
	\begin{pmatrix}
		\widetilde{\mathbf{V}}_Q \\
		\widetilde{\mathbf{V}}_R
	\end{pmatrix}
	=
	\begin{pmatrix}
		\widetilde{\mathbf{A}} \\
		\widetilde{\mathbf{B}}
	\end{pmatrix},
\end{equation}
where $\kappa = (\zeta_1+\zeta_2)/(\zeta_1\zeta_2)H$, $\widetilde{\mathbf V}_Q$ and $\widetilde{\mathbf V}_R$ denote the translational and internal velocities in Fourier space, and $\widetilde{\mathbf A}$ and $\widetilde{\mathbf B}$ are the corresponding noise terms.

Solving the linear system gives
\begin{equation}
\widetilde{\mathbf{V}}_Q = \frac{d}{\Delta}\widetilde{\mathbf{A}} - \frac{i\omega M_{12}}{\Delta}\widetilde{\mathbf{B}},
\end{equation}
with $d = i\omega M_{22}+1+\frac{\kappa}{i\omega}$, $\Delta = (i\omega M_{11}+1)d + \omega^2 M_{12}M_{21}$.

The fluctuation--dissipation theorem yields
\begin{equation}
\begin{aligned}
\langle \widetilde{A}_\alpha(\omega) \widetilde{A}_\beta^*(\omega') \rangle
&= 2\pi S_{AA}(\omega)\delta_{\alpha\beta}\delta(\omega-\omega'), \\
\langle \widetilde{B}_\alpha(\omega) \widetilde{B}_\beta^*(\omega') \rangle
&= 2\pi S_{BB}(\omega)\delta_{\alpha\beta}\delta(\omega-\omega').
\end{aligned}
\end{equation}
with $S_{AA}(\omega)= \frac{2k_B T}{\zeta_1+\zeta_2}$, $S_{BB}(\omega)= 2k_B T \frac{\zeta_1+\zeta_2}{\zeta_1\zeta_2}$ 
and $\langle \widetilde{A}_\alpha(\omega) \widetilde{B}_\beta^*(\omega') \rangle = 0$.
It then follows that
\begin{equation}
S_{\dot Q\dot Q}(\omega)
=
\left|\frac{d}{\Delta}\right|^2 S_{AA}(\omega)
+
\left|\frac{i\omega M_{12}}{\Delta}\right|^2 S_{BB}(\omega).
\end{equation}
The translational diffusivity is obtained from the Green--Kubo relation $D_Q = \frac{1}{2} S_{\dot{Q}\dot{Q}}(0)$, which is the standard equilibrium linear-response expression for diffusion in Langevin systems.\cite{zwanzig2001nonequilibrium,gao2024some,malaspina2024green} 
Taking the limit $\omega \to 0$ gives
\begin{equation}
S_{\dot{Q}\dot{Q}}(0) = S_{AA}(0) = \frac{2k_B T}{\zeta_1+\zeta_2},
\end{equation}
hence
\begin{equation}\label{eq:D_Q}
D_Q = \frac{k_B T}{\zeta_1+\zeta_2}.
\end{equation}
This result coincides with the overdamped prediction. Thus, inertia and the $\mathbf Q$--$\mathbf R$ coupling affect only the short-time dynamics, while the long-time translational diffusivity remains unchanged.

\begin{figure}[htbh]\centering
\includegraphics[width=0.85\linewidth]{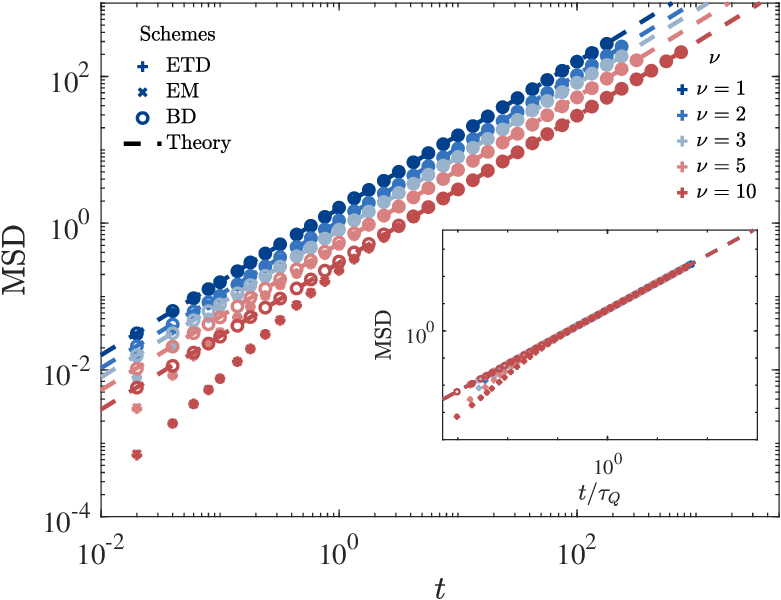}
\caption{Mean-squared displacement (MSD) of the center-of-resistance $\mathbf Q$ versus the physical time $t$ for different asymmetry ratios $\nu$. The inset replots the same data versus the scaled time $t/\tau_Q$, where $\tau_Q\equiv(\zeta_1+\zeta_2)/H$ is an empirical crossover timescale that collapses the onset of the long-time diffusive regime. 
The dashed line indicates the asymptotic diffusion law of Eq.~\eqref{eq:D_Q}.
Statistical uncertainties are smaller than the plotting symbols.}
\label{fig.dumbbellmsdcor}
\end{figure}

The translational properties can be further examined in the time domain through the mean-squared displacement (MSD) of the center of resistance, $\langle \Delta Q^2(t)\rangle$. Fig.~\ref{fig.dumbbellmsdcor} shows the MSD obtained from the ETD, EM, and BD simulations for different asymmetry ratios $\nu$.
At long times, all curves collapse onto a linear scaling,
\begin{equation}
	\langle \Delta Q^2(t)\rangle \simeq 6D_Q t,
\end{equation}
with the diffusion coefficient ($D_Q$) in agreement with the theoretical prediction derived above. This confirms that neither inertia nor the coupling between the translational and internal modes modifies the long-time transport behavior.
A clear difference appears at short times. While the BD data reaches the diffusive regime rapidly, the EM and ETD results fall below it for large $\nu$. The deviation increases with $\nu$ and extends over a broader time range, but eventually vanishes as the dynamics cross over to the common long-time diffusive regime. The agreement between EM and ETD further indicates that this trend is physical rather than a numerical artifact.

\begin{figure*}[htbp]
\centering
\begin{subfigure}[t]{0.45\textwidth} 
\centering
\includegraphics[height=0.25\textheight]{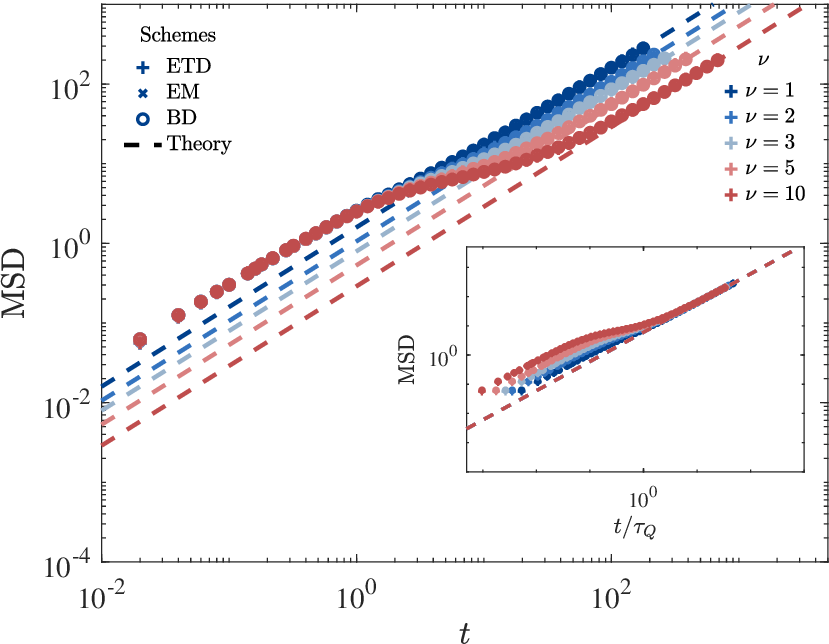}
\caption{$a_1$}
\end{subfigure}
\begin{subfigure}[t]{0.45\textwidth} 
\centering
\includegraphics[height=0.25\textheight]{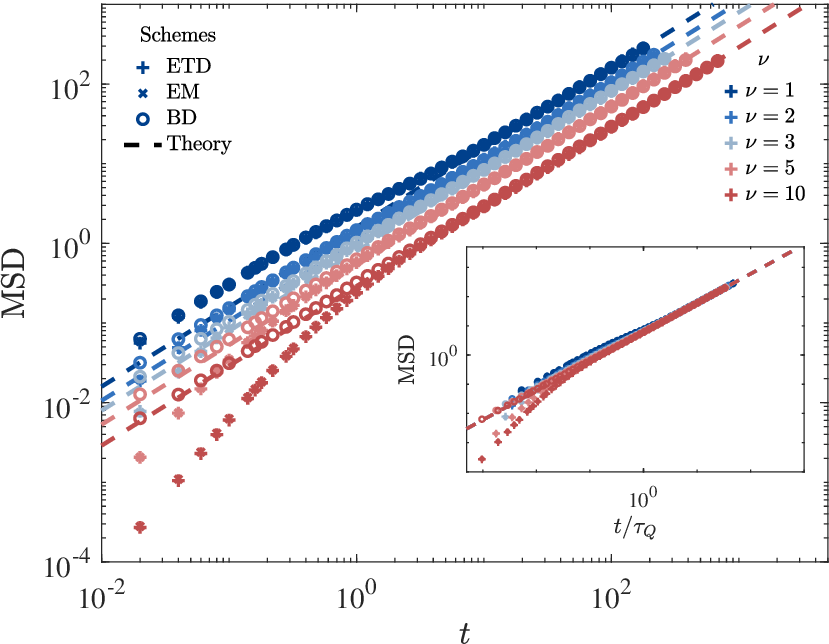}
\caption{$a_2=\nu a_1$}
\end{subfigure}

\par 

\begin{subfigure}[t]{0.45\textwidth} 
\centering
\includegraphics[height=0.25\textheight]{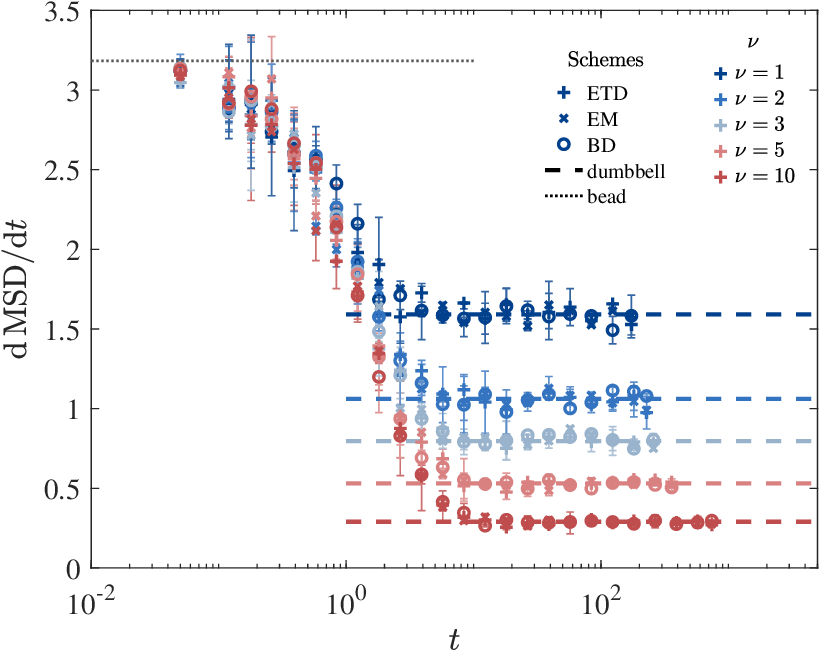}
\caption{$a_1$}
\end{subfigure}
\begin{subfigure}[t]{0.45\textwidth} 
\centering
\includegraphics[height=0.25\textheight]{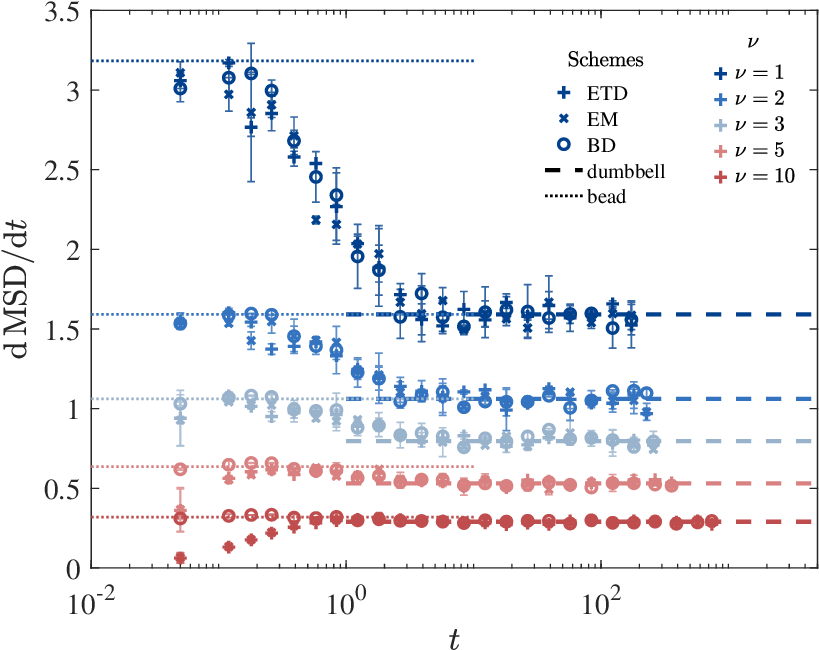}
\caption{$a_2=\nu a_1$}
\end{subfigure}

\caption{
Mean-squared displacements (MSDs) of the two beads and their time derivatives for different asymmetry ratios $\nu$. 
Panels (a) and (c) correspond to bead 1 of fixed radius $a_1$, while panels (b) and (d) correspond to bead 2 with radius $a_2=\nu a_1$. 
The upper panels show the MSD as a function of $t$, and the insets display the same data versus the scaled time $t/\tau_Q$. 
In the lower panels, dashed horizontal lines indicate the long-time diffusivity of the dumbbell, whereas dotted horizontal lines indicate the diffusivity of the corresponding isolated bead.
{Error bars in the lower panels indicate statistical uncertainties in the numerical derivative.}}
\label{fig.dumbbellmsdbeads}
\end{figure*}
To clarify how the transient translational response is partitioned between the two beads, we further examine the bead-resolved MSDs and their time derivatives in Fig.~\ref{fig.dumbbellmsdbeads}. 
{Statistical uncertainties for the MSD are smaller than the symbol size across the entire time range and are therefore not shown.}
Here, bead 1 has fixed radius $a_1$, whereas bead 2 has radius $a_2=\nu a_1$. In all cases, the long-time MSDs of the two beads recover the same linear growth, confirming that both beads ultimately share the translational diffusion of the dumbbell. The effect of asymmetry is therefore confined to the transient regime.
The derivative plots reveal two characteristic plateau values, corresponding to the diffusivity of the isolated bead and the long-time diffusivity of the dumbbell. 
{Error bars indicate statistical uncertainties estimated from independent trajectory samples.}
Bead 1 remains close to overdamped behavior: its dynamics begins near the isolated-bead plateau and then crosses over to the dumbbell plateau, with weak dependence on $\nu$. By contrast, for moderate asymmetry ratios, bead 2 exhibits a pronounced four-stage evolution: an initial inertial build-up from nearly zero diffusivity, an intermediate isolated-bead plateau, a crossover regime, and the final dumbbell plateau.
The internal configurational relaxation time $\tau_R = \zeta_1\zeta_2/[(\zeta_1+\zeta_2)H]$ marks the onset of the final crossover, namely the departure from the isolated-bead plateau. The full establishment of dumbbell diffusion occurs only at a later time, since the crossover extends over a finite interval. As $\nu$ increases, the onset of this final crossover is observed at later times, consistent with the growth of $\tau_R$, while the isolated-bead and dumbbell diffusivities become closer, making the intermediate plateau progressively shorter and less distinct. Finite mass does not alter the asymptotic translational diffusivity of the dumbbell, but it redistributes the transient mobility unevenly between the two beads: the smaller bead remains nearly overdamped, whereas the larger bead carries the dominant inertial and asymmetry-dependent correction.

\begin{figure*}[htbp]
	\centering
	\begin{subfigure}[b]{0.48\textwidth} 
		\centering
		\includegraphics[width=0.85\linewidth]{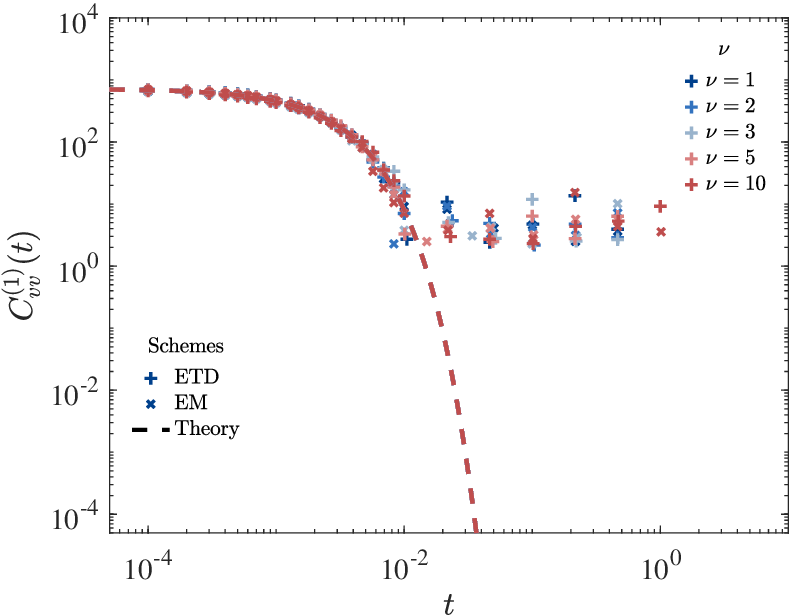}
		\caption{$a_1$}
	\end{subfigure}
	\begin{subfigure}[b]{0.48\textwidth} 
		\centering
		\includegraphics[width=0.85\linewidth]{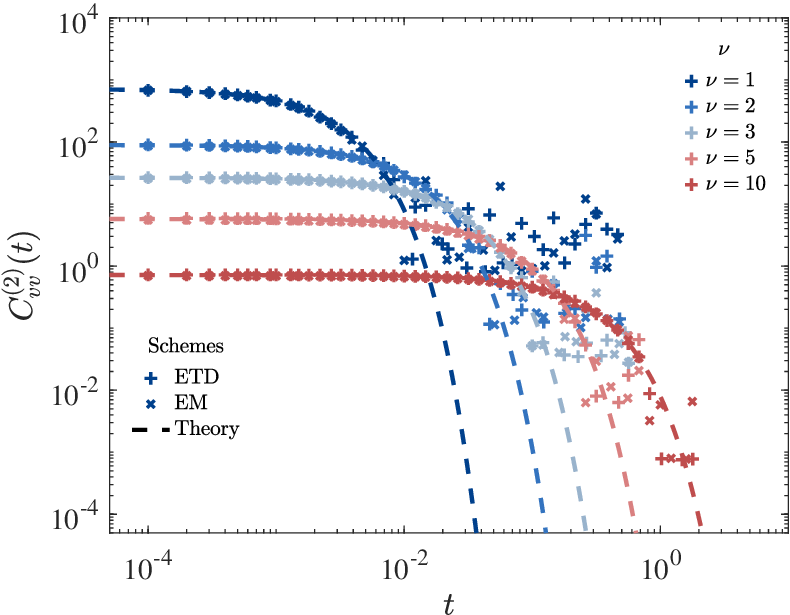}
		\caption{$a_2=\nu a_1$}
	\end{subfigure}
    \caption{
    Velocity autocorrelation functions (VACFs) of the two beads in a lopsided dumbbell for several asymmetry ratios $\nu$, plotted against time $t$. Bead 1 has fixed radius $a_1$, while bead 2 has radius $a_2=\nu a_1$.
    (a) VACF of bead 1.
    (b) VACF of bead 2.
    Dashed lines show the short-time single-bead prediction from Eq.~\eqref{eq.single_bead_vacf}.}
	\label{fig.dumbbellbeadsvacf}
\end{figure*}
\paragraph*{Short-time velocity correlations.}
To further verify the short-time inertial dynamics, we examine the bead-resolved velocity autocorrelation functions (VACFs) in Fig.~\ref{fig.dumbbellbeadsvacf}. In the short-time regime, the numerical results for both beads follow closely the corresponding single-bead Langevin prediction,\cite{langevin1908theorie,uhlenbeck1930theory,Dingyi2025}
\begin{equation}
	C_{vv}^{(i)}(t)=\frac{3k_BT}{m_i}e^{-t/\tau_i},
	\quad
	\tau_i=\frac{m_i}{\zeta_i},
    \label{eq.single_bead_vacf}
\end{equation}
showing that the initial decay of velocity memory is governed primarily by the local bead properties. The agreement confirms that, before the spring coupling becomes significant, each bead relaxes essentially as an individual Langevin particle.

\begin{figure}[tbh]\centering
	\includegraphics[width=0.85\linewidth]{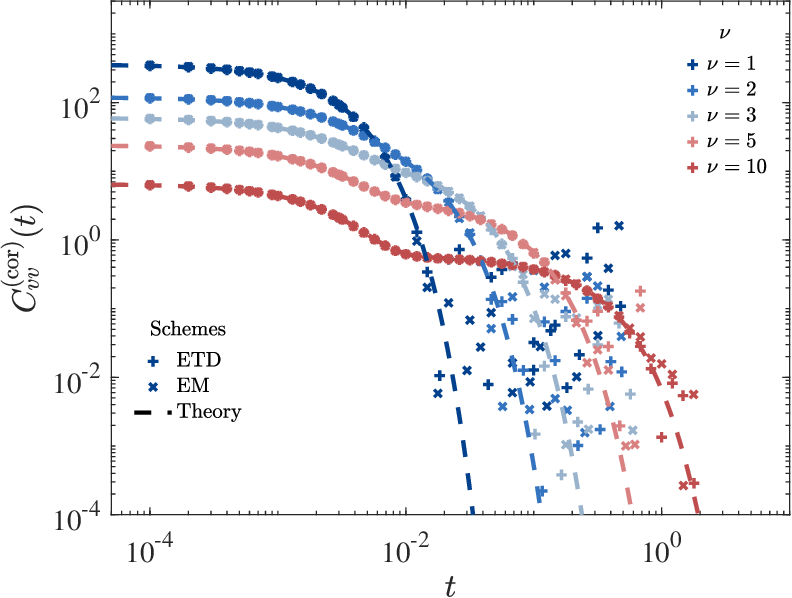}
    \caption{Velocity autocorrelation function (VACF) of the center of resistance, $C_{vv}^{(\mathrm{cor})}(t)$, plotted against time $t$ for different asymmetry ratios $\nu$. 
    Dashed lines show the short-time theoretical prediction given by a weighted sum of two exponential relaxations.}
	\label{fig.dumbbellvacfcor}
\end{figure}
The short-time VACF of the center of resistance is likewise well described by a weighted superposition of the two single-bead inertial relaxations, obtained by neglecting inter-bead coupling at short times,
\begin{equation}
    C_{vv}^{(\mathrm{cor})}(t) =\frac{3k_BT}{(\zeta_1+\zeta_2)^2}
    \left[W_1 e^{-t/\tau_{m,1}}+W_2 e^{-t/\tau_{m,2}}\right],
\end{equation}
where $\tau_{m,i}=m_i/\zeta_i$ and $W_i=\zeta_i^2/m_i$. As shown in Fig.~\ref{fig.dumbbellvacfcor}, the numerical results agree well with this weighted two-exponential form over the resolved short-time range. This confirms that the short-time center-of-resistance dynamics is determined by the superposition of the two local inertial relaxation modes.

\subsection{Configuration Dynamics}
\label{sec:config}
\paragraph*{Equilibrium statistics and spectral response.}
We now turn to the internal dynamics of the dumbbell, characterized by the end-to-end vector $\mathbf{R}=\mathbf{r}_2-\mathbf{r}_1$. In the classical overdamped dumbbell theory, obtained after eliminating the momentum variables from the Langevin description, the relative coordinate follows an Ornstein--Uhlenbeck process, implying a Gaussian equilibrium distribution and closed-form predictions for the linear correlators of $\mathbf{R}$. \cite{uhlenbeck1930theory,doi1988theory,bird1987dynamics} We quantify how finite inertia and the resulting $\mathbf{Q}$--$\mathbf{R}$ coupling modify these overdamped predictions.
Our BD simulations (strictly overdamped) provide a numerical benchmark for the overdamped theory. The ETD and EM simulations include inertia and thus quantify the deviations induced by finite mass and by the $\mathbf{Q}$--$\mathbf{R}$ coupling. In the following, we compare all simulations against the overdamped predictions and identify the time window over which inertial effects are observable.

\begin{figure}[tbh]\centering
	\includegraphics[width=0.85\linewidth]{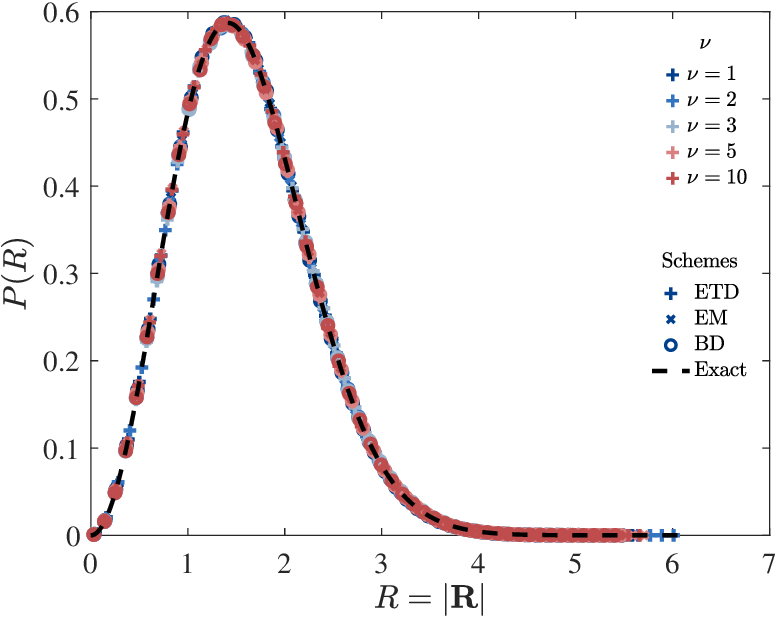}
	\caption{Equilibrium probability density of the end-to-end distance $R=|\mathbf{R}|$ for several asymmetry ratios $\nu$. 
    The dashed curve is the Boltzmann prediction for a 3D Hookean spring, Eq.~\eqref{eq.dumbbellPR}.
    }
	\label{fig.dumbbellPR}
\end{figure}

We first check that all integrators sample the correct equilibrium statistics. For a Hookean spring,
$U(R)=\tfrac12 H R^2$, equilibrium depends only on $U$ and $k_BT$, and is therefore independent of the bead masses and friction coefficients. The exact distribution of $R=|\mathbf{R}|$ follows from integrating the Boltzmann factor over angles,
\begin{equation}
	P(R)=4\pi\left(\frac{H}{2\pi k_B T}\right)^{3/2} R^2
	\exp\!\left(-\frac{H R^2}{2k_B T}\right),
	\label{eq.dumbbellPR}
\end{equation}
and the Cartesian components are Gaussian with $\langle R^2\rangle=3k_BT/H$.
Figure~\ref{fig.dumbbellPR} shows that ETD, EM, and BD all recover Eq.~\eqref{eq.dumbbellPR} for every $\nu$ within statistical uncertainty, indicating that the underdamped schemes preserve the correct Boltzmann statistics for the configurational degrees of freedom.

\paragraph*{Power spectral density.}
The dumbbell features two distinct friction scales associated with different modes. The translational mode (e.g., the center-of-resistance $\mathbf Q$) experiences the additive drag $\zeta_{\rm trans}=\zeta_1+\zeta_2$, yielding $D_Q=k_BT/(\zeta_1+\zeta_2)$. In contrast, the internal mode $\mathbf R=\mathbf r_2-\mathbf r_1$ relaxes with the effective friction $\zeta_{\rm eff}=\zeta_1\zeta_2/(\zeta_1+\zeta_2)$, so that for a Hookean spring we characterize configurational relaxation by~\cite{doi1988theory,phan2024lopsided}
\begin{equation}
\tau_R \equiv \frac{\zeta_{\mathrm{eff}}}{H},\quad 
\frac{1}{\zeta_{\mathrm{eff}}}=\frac{1}{\zeta_1}+\frac{1}{\zeta_2},
\end{equation}
which is the natural timescale for the relative coordinate in the overdamped theory. In the time domain, this scale controls the dominant configurational relaxation, whereas finite inertia is expected to appear most clearly in the short-time and high-frequency response. It is therefore useful to examine the power spectral density (PSD) of the end-to-end vector. For convenience, we first work with the two-sided spectrum in angular frequency $\omega$ and then convert to the one-sided PSD reported by the FFT in terms of the ordinary frequency $f\ge 0$. Within the present linear framework, the spectrum follows from the Fourier-transformed coupled equations in Eq.~\eqref{eq.dumbbelllangevin}. Solving for the internal velocity $\widetilde{\mathbf V}_R(\omega)$ gives
\begin{equation}
	\widetilde{\mathbf V}_R
	=
	-\frac{i\omega M_{21}}{\Delta}\widetilde{\mathbf A}
	+
	\frac{i\omega M_{11}+1}{\Delta}\widetilde{\mathbf B},
\end{equation}
and, using $\widetilde{\mathbf R}=\widetilde{\mathbf V}_R/(i\omega)$,
\begin{equation}
	\widetilde{\mathbf R}
	=
	-\frac{M_{21}}{\Delta}\widetilde{\mathbf A}
	+
	\frac{M_{11}+(i\omega)^{-1}}{\Delta}\widetilde{\mathbf B}.
\end{equation}
Because the stochastic forcings $\widetilde{\mathbf A}$ and $\widetilde{\mathbf B}$ are uncorrelated, the two-sided angular-frequency PSD of $\mathbf R$, summed over the three Cartesian directions, is
\begin{equation}
\begin{aligned}
    S_{RR,\omega}(\omega)
	&=
	3\left|\frac{M_{21}}{\Delta}\right|^2 S_{AA}(\omega)
	\\
    &+
	3\left|\frac{M_{11}+(i\omega)^{-1}}{\Delta}\right|^2 S_{BB}(\omega).
\end{aligned}
\label{eq.exact_SRR_omega}
\end{equation}
To compare with the numerical FFT output, we introduce the one-sided PSD in ordinary frequency, defined by
\begin{equation}
	\langle R^2\rangle = \int_0^\infty S_{RR}(f)\,df ,
\end{equation}
which is related to the two-sided angular-frequency spectrum through
\begin{equation}
	S_{RR}(f)=4\pi\,S_{RR,\omega}(2\pi f), \qquad f>0.
	\label{eq.one_sided_conversion}
\end{equation}
All spectra shown below are presented in this one-sided frequency convention, and we therefore use $S_{RR}(f)$ as the PSD notation hereafter.
Equation~\eqref{eq.exact_SRR_omega} makes the overdamped and inertial limits transparent; below, we quote the intermediate angular-frequency expressions only when needed for the conversion to the one-sided PSD $S_{RR}(f)$.
In the BD limit $M_{ij}\to0$, $\Delta\to 1+\kappa/(i\omega)$ and Eq.~\eqref{eq.exact_SRR_omega} reduces to
\begin{equation}
	S_{RR,\omega}^{BD}(\omega)
	=
	\frac{3S_{BB}(\omega)}{\omega^2+\kappa^2}.
\end{equation}
Using $S_{BB}(\omega)=2k_B T(\zeta_1+\zeta_2)/(\zeta_1\zeta_2)=2k_B T\,\kappa/H$ and $\tau_R=\kappa^{-1}$, one obtains
\begin{equation}
	S_{RR,\omega}^{BD}(\omega)
	=
	\frac{6k_B T}{H}\frac{\tau_R}{1+(\omega\tau_R)^2},
\end{equation}
or equivalently, in the one-sided frequency representation,
\begin{equation}
\begin{aligned}
    S_{RR}^{BD}(f)
	&=
	4\pi S_{RR,\omega}^{BD}(2\pi f)
	\\
    &=
	\frac{24\pi k_B T}{H}
	\frac{\tau_R}{1+(2\pi f\tau_R)^2}.
\end{aligned}\label{eq.SRR_BD_one_sided}
\end{equation}
Thus, the overdamped one-sided PSD decays as $S_{RR}^{BD}(f)\sim f^{-2}$ at high frequency.

For finite bead masses, the high-frequency behavior is modified by the inertial matrix. As $\omega \to \infty$, the $i\omega M$ terms dominate the coefficients, so
\begin{equation}
	\Delta \sim -\omega^2(M_{11}M_{22}-M_{12}M_{21})=-\omega^2\det \left( {\bf M} \right).
\end{equation}
Here $\det \left( {\bf M} \right)$ is the determinant of the inertial matrix introduced above, with $\det \left( {\bf M} \right)=m_1m_2/(\zeta_1\zeta_2)$.
Substituting these asymptotic forms into Eq.~\eqref{eq.exact_SRR_omega} yields
\begin{equation}
\begin{aligned}
   	S_{RR,\omega}^{\mathrm{inertial}}&(\omega)
	\simeq
    \\
	&\frac{3M_{21}^2S_{AA}(\omega)+3M_{11}^2S_{BB}(\omega)}{\omega^4[\det \left( {\bf M} \right)]^2},  \quad\omega\to\infty.
\end{aligned}
\end{equation}
This asymptotic scaling reflects the dominance of the inertial operator in the high-frequency limit, rather than discretization effects. 
Accordingly, inertial effects are expected on the relevant momentum-relaxation timescale $t\sim\tau_m$, where $\tau_m$ denotes the characteristic bead-scale momentum relaxation time, or equivalently for frequencies $\omega \gtrsim 1/\tau_m$.
In this regime, the inertial terms dominate the viscous and elastic contributions, leading to the high-frequency scaling $S_{RR}(f)\sim f^{-4}$.
In practice, the BD reference spectrum extracted from simulation is affected by discrete sampling. For a trajectory sampled at time interval $h$, each Cartesian component of the overdamped connector vector follows a discretely sampled Ornstein--Uhlenbeck process~\cite{uhlenbeck1930theory} with one-sided PSD
\begin{equation}
\begin{aligned}
  	S_{RR}^{\mathrm{sampled\mbox{-}BD}}(f)
	&=
    \\
	&\frac{24\pi h\,k_B T}{H}
	\frac{1-a^2}{1+a^2-2a\cos(2\pi fh)},
\end{aligned}\label{eq.sampled_BD_PSD}
\end{equation}
where $a=\exp(-h/\tau_R)$ and the factor of $3$ from the three Cartesian directions has already been included. Equation~\eqref{eq.sampled_BD_PSD} reduces to Eq.~\eqref{eq.SRR_BD_one_sided} in the continuous-sampling limit $h/\tau_R\to 0$.
\begin{figure}[tbh]\centering
	\includegraphics[width=0.85\linewidth]{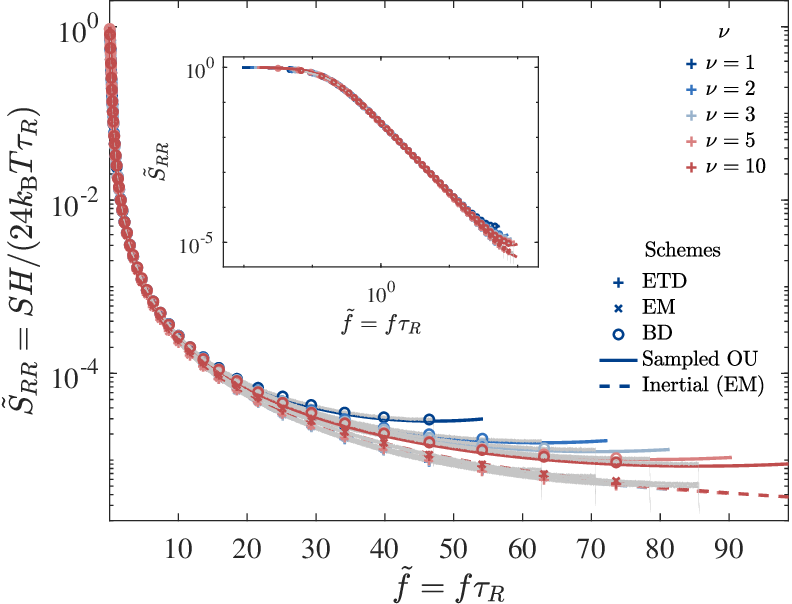}
	\caption{Dimensionless one-sided power spectral density of the end-to-end vector, $\widetilde S_{RR}$ 
    plotted against the dimensionless frequency $\widetilde f$ 
    for varying asymmetry ratios $\nu$. 
    The solid line represents the continuous-time overdamped prediction, while the dashed line shows the discrete-time sampled-BD spectrum in Eq.~\eqref{eq.sampled_BD_PSD}. The inset enlarges the low-frequency region.
    {Shaded bands indicate statistical uncertainties.}}
	\label{fig.dumbbellPSDR}
\end{figure}
The dimensionless PSD, $\widetilde S_{RR}=S_{RR}H/(24\pi k_B T\tau_R)$, is plotted against the scaled frequency $\widetilde f=f\tau_R$ in Fig.~\ref{fig.dumbbellPSDR}. 
{Statistical uncertainties are indicated by the shaded confidence bands and remain comparable to or smaller than the symbol size, and are thus negligible for the subsequent analysis.}
In the low-frequency regime, $\widetilde f\lesssim 1$, the spectra for different $\nu$ collapse onto a common plateau, indicating that the long-time configurational relaxation is governed primarily by the overdamped spring mode and depends only weakly on the asymmetry ratio. In the intermediate-frequency range, the BD results are well captured by the sampled-BD reference, confirming that the small deviations from the continuous Lorentzian line arise mainly from discrete sampling. At higher frequencies, the ETD and EM spectra remain close to each other but fall below the overdamped BD prediction. This suppression of the high-frequency tail is consistent with the inertial asymptotics derived above: mass and friction asymmetry couple momentum and configurational relaxation, thereby accelerating the decay of short-time structural fluctuations.
This indicates that inertia mainly modifies the shortest-time configurational dynamics, while the dominant low-frequency relaxation remains close to the overdamped description. The spectrum accordingly exhibits three frequency regimes: at low frequencies, $f \ll 1/\tau_R$, it is dominated by configurational relaxation and remains close to the overdamped prediction; at intermediate frequencies, $1/\tau_R \ll f \ll 1/\tau_m$, the overdamped scaling still provides an approximate description, although discrete-sampling and finite-inertia effects begin to appear; and at high frequencies, $f \gg 1/\tau_m$, the inertial asymptotics dominate and enforce the steeper $f^{-4}$ decay.

\paragraph*{Time-domain correlations.}
The same configurational relaxation can be characterized in the time domain through the autocorrelations
\begin{equation}
\begin{aligned}
    C_{RR}(t)&=\langle R^2\rangle e^{-t/\tau_R},\\
    C_{R^2R^2}(t)&=2\langle R^2\rangle^2 e^{-2t/\tau_R},
\end{aligned}
\end{equation}
which provide the natural overdamped reference in the time domain.
\begin{figure*}[htbp]
	\centering
	\begin{subfigure}[b]{0.48\textwidth} 
		\centering
		\includegraphics[width=0.85\linewidth]{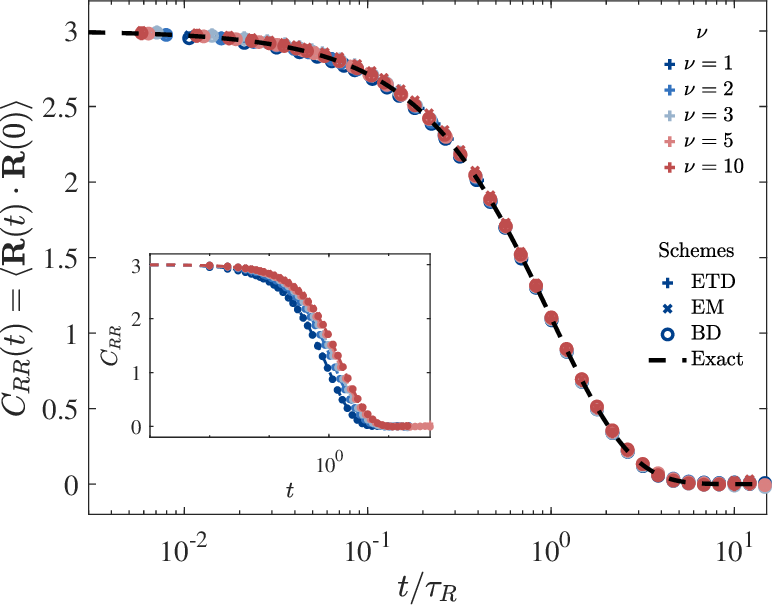}
		\caption{}
	\end{subfigure}
	\begin{subfigure}[b]{0.48\textwidth} 
		\centering
		\includegraphics[width=0.85\linewidth]{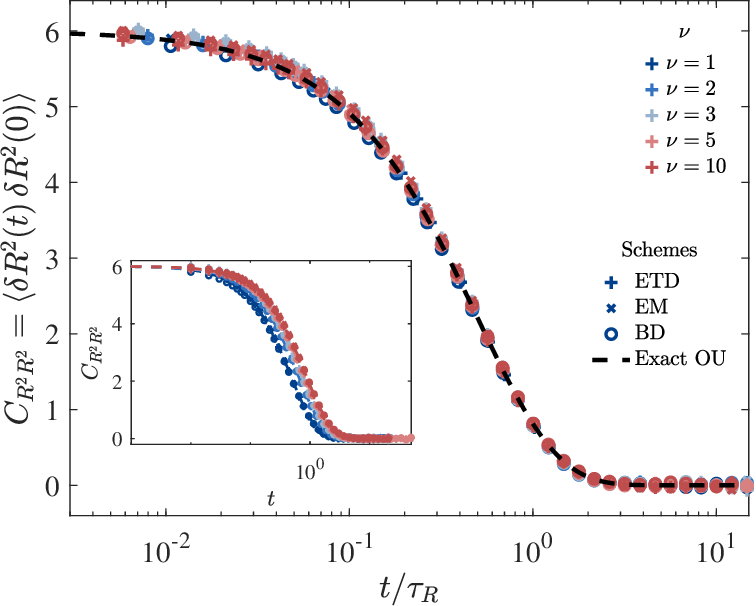}
		\caption{}
	\end{subfigure}
	\caption{Configurational autocorrelations of lopsided elastic dumbbells for several asymmetry ratios $\nu$: (a) $C_{RR}(t)$ and (b) $C_{R^2R^2}(t)$. 
    Dashed lines show the overdamped predictions. 
    }
	\label{fig.dumbbellRRR2R2}
\end{figure*}
As shown in Fig.~\ref{fig.dumbbellRRR2R2}, scaling time by $\tau_R$ collapses the data for all tested asymmetry ratios $\nu$ onto the overdamped master curves. Over the resolved time range, the underdamped integrators (EM and ETD) are indistinguishable from the overdamped BD results and from the analytical predictions, indicating that inertial effects are confined to an initial transient that is not visible on the scale of $\tau_R$.
To quantify this separation of timescales, we compare $\tau_R$ with the internal momentum relaxation time $\tau_m$. Using the asymmetry ratio $\nu=a_2/a_1$ and the scalings $m_2=\nu^3 m_1$ and $\zeta_2=\nu \zeta_1$, one finds 
\[
\tau_R=\frac{\nu}{\nu+1}\frac{\zeta_1}{H},
\]
and 
\[
\tau_m=\frac{m_1+m_2}{\zeta_{\mathrm{eff}}}
=\frac{\nu^2}{\nu^2-\nu+1}\frac{m_1}{\zeta_1}.
\]
In our simulations $\tau_m(\nu)\ll \tau_R(\nu)$ for all tested $\nu$, so any inertial signature is restricted to very short times and is overwhelmed by the subsequent exponential configurational relaxation. Consequently, the overdamped approximation remains accurate for the low-frequency structural dynamics even when inertia is retained in the integrator.

\begin{figure}[tbh]\centering
	\includegraphics[width=0.85\linewidth]{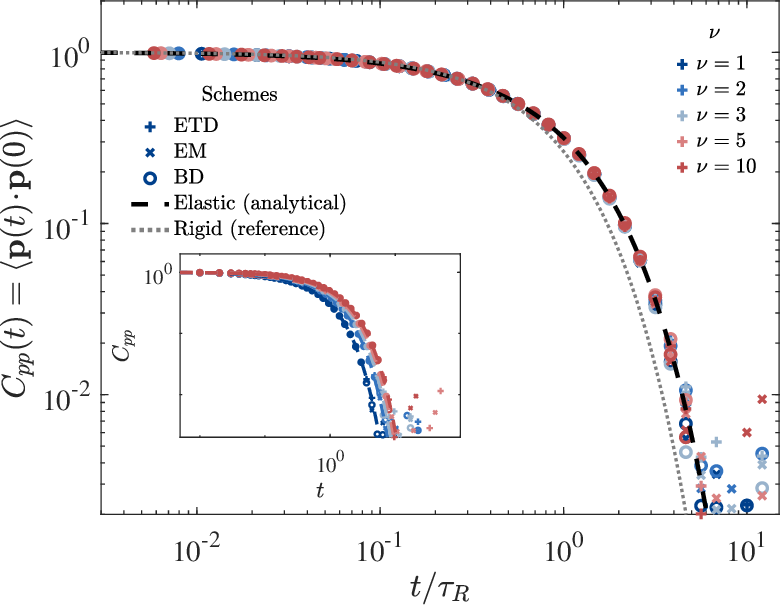}
	\caption{Orientational autocorrelation $C_{pp}(t)$ 
    for lopsided elastic dumbbells at different asymmetry ratios $\nu$. 
    Dashed line is the analytical prediction in Eq.~\eqref{eq.exact_Cuu}, 
    {while the dotted gray line denotes the rigid dumbbell reference based on Eq.~\eqref{eq:rigiddumbbellrotatory}.}
    }
	\label{fig.dumbbellCuu}
\end{figure}
This long-time insensitivity to finite mass extends to the orientational relaxation, quantified by the autocorrelation  function 
\begin{equation}
    C_{pp}(t)=\langle \mathbf p(t)\!\cdot\!\mathbf p(0)\rangle,
\end{equation} 
where $\mathbf p = \mathbf R/|\mathbf R|$. 
As a reference, in the rigid-dumbbell limit ($|\mathbf R|=L$ fixed), the orientation undergoes pure rotational diffusion with rotational diffusivity 
\begin{equation}
D_r=\frac{2k_BT(\zeta_1+\zeta_2)}{\zeta_1\zeta_2 L^2},
\label{eq:rigiddumbbellrotatory}
\end{equation}
following the standard rigid dumbbell theory.~\cite{bird1987dynamics,phan2024lopsidedtutorial} The corresponding orientational autocorrelation therefore takes the standard first-rank rotational-diffusion form 
\[
C_{pp}(t)=\langle \mathbf p(t)\!\cdot\!\mathbf p(0)\rangle=e^{-2D_r t}.
\]
For an elastic dumbbell, the extension $R(t)$ fluctuates, so orientational relaxation is coupled to internal configurational dynamics and is not, in general, captured by a single-exponential form with a constant $D_r$. 
In the Hookean case, $\mathbf R(t)$ remains a linear Ornstein--Uhlenbeck process, so $C_{pp}(t)$ can be obtained in closed form. We use this overdamped result as a baseline and then examine how finite inertia in the lopsided dynamics alters the short-time behavior.
Specifically, in the overdamped Hookean case, $\mathbf R(t)$ is a stationary, isotropic Gaussian Markov process and admits the standard Ornstein--Uhlenbeck representation~\cite{uhlenbeck1930theory,doi1988theory}
\begin{equation}
	\mathbf{R}(t)=\rho(t)\,\mathbf{R}(0)+\sqrt{1-\rho^2(t)}\,\mathbf{Y}(t),
	\label{eq.R_decomposition}
\end{equation}
where $\rho(t)=\exp(-t/\tau_R)$ and $\mathbf{Y}(t)$ is an independent isotropic Gaussian vector with the same covariance as $\mathbf{R}(0)$. Let $X=|\mathbf R(0)|$, $Y=|\mathbf Y(t)|$, and
$w=\mathbf{R}(0)\!\cdot\!\mathbf{Y}(t)/(XY)$. By isotropy, $w$ is uniformly distributed on $[-1,1]$, while $X$ and $Y$ follow the same Maxwell distribution. Equation~\eqref{eq.R_decomposition} then gives
\begin{equation}
\begin{aligned}
&C_{pp}(t)=
\\
&\Biggl\langle
\frac{\rho(t)X+\sqrt{1-\rho^2(t)}\,Yw}
{\sqrt{\rho^2(t)X^2+(1-\rho^2(t))Y^2+2\rho(t)\sqrt{1-\rho^2(t)}\,XYw}}
\Biggr\rangle,
\label{eq:Cpp_integral_form}
\end{aligned}
\end{equation}
where the average is taken over the distributions of $X$, $Y$, and $w$. Carrying out these integrations yields, to the best of our knowledge, the following closed-form result for the overdamped Hookean dumbbell:
\begin{equation}
\begin{aligned}
	&C_{pp}(t) = 
    \\
    &\frac{2}{\pi \rho^2(t)}\left[(2\rho^2(t)-1)\arcsin\rho(t)+\rho(t)\sqrt{1-\rho^2(t)}\right].
\end{aligned}\label{eq.exact_Cuu}
\end{equation}
Eq.~\eqref{eq.exact_Cuu} provides an overdamped Hookean reference for the orientational relaxation. As shown in Fig.~\ref{fig.dumbbellCuu}, $C_{pp}(t)$ obtained from the underdamped integrators (EM and ETD) agrees with the overdamped BD results and collapses onto Eq.~\eqref{eq.exact_Cuu} when time is rescaled by $\tau_R$, for all tested asymmetry ratios $\nu$. Although the underdamped dynamics enforces $C'_{pp}(0)=0$ on the momentum-relaxation scale $t=O(\tau_m)$, this $t\!\to\!0$ constraint does not affect the relaxation on the configurational timescale $t=O(\tau_R)$ resolved in Fig.~\ref{fig.dumbbellCuu}.
{For reference, we also include the rigid dumbbell prediction based on the rotational diffusion coefficient in Eq.~\eqref{eq:rigiddumbbellrotatory}. The rigid and elastic dumbbell models are fundamentally different: the rigid model enforces a fixed connector length, whereas in the elastic case the connector undergoes thermal fluctuations and is coupled to the orientational dynamics. Accordingly, the rigid dumbbell serves only as a limiting reference, and a direct numerical comparison based on rigid dumbbell simulations is beyond the scope of the present work.}

\begin{figure}[tbh]\centering
	\includegraphics[width=0.85\linewidth]{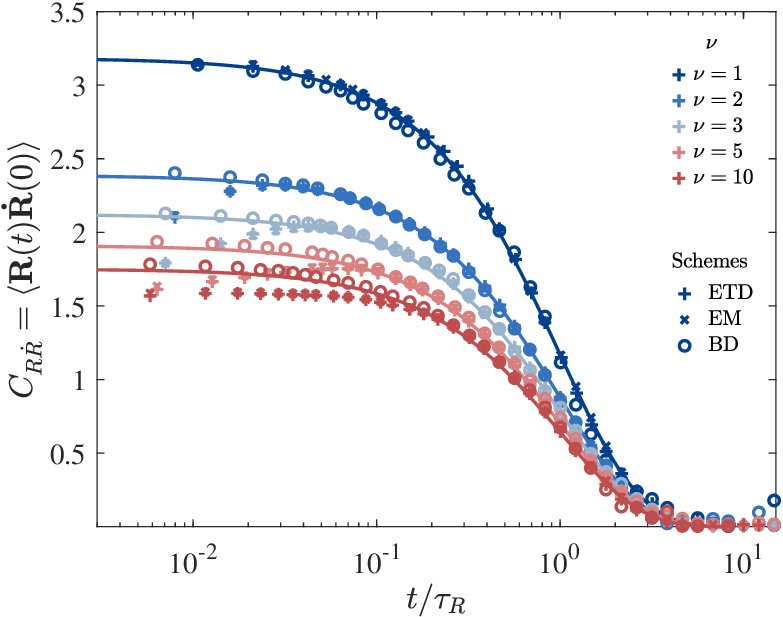}
	\caption{Position--velocity cross-correlation $C_{R\dot{R}}(t)$ 
    versus scaled time $t/\tau_R$ for several asymmetry ratios $\nu$. 
    Solid lines are overdamped prediction by Eq.~\eqref{eq:c_RRdot}.
    }
	\label{fig.dumbbellCRV}
\end{figure}
In addition to the spectral analysis, the transient coupling between configurational and velocity fluctuations can be examined in the time domain through the position--velocity cross-correlation function
\begin{equation}
    C_{R\dot{R}}(t)=\langle \mathbf{R}(0)\cdot\dot{\mathbf{R}}(t)\rangle.
\end{equation}
In the overdamped limit, this quantity is expected to follow
\begin{equation}\label{eq:c_RRdot}
	C_{R\dot{R}}(t)
	=
	-\frac{3k_BT}{\zeta_{\mathrm{eff}}}e^{-t/\tau_R},
	\qquad t>0,
\end{equation}
which provides the reference relaxation curve for the configurational mode.
Fig.~\ref{fig.dumbbellCRV} shows $C_{R\dot{R}}(t)$ as a function of the scaled time $t/\tau_R$ for different asymmetry ratios $\nu$. The BD data follow the theoretical curves closely over the full time range, confirming the overdamped prediction for this observable. The EM and ETD results also remain nearly indistinguishable from each other, indicating that the observed deviations are not caused by the choice of numerical integrator.
A systematic departure appears, however, when $\nu>1$. In this regime, the EM and ETD data lie below the overdamped prediction at short times and recover the BD curve only at later times. By contrast, for $\nu=1$, the underdamped and overdamped results remain in close agreement over the full time range. These results indicate that inertia does not substantially alter the long-time configurational relaxation, but introduces a transient short-time suppression of the position--velocity correlation when the mass asymmetry is finite. Moreover, the crossover time beyond which the EM/ETD results merge with the overdamped prediction increases with $\nu$, showing that the duration of this inertial transient becomes longer as the asymmetry ratio $\nu$ increases.

\begin{figure}[tbh]\centering
	\includegraphics[width=0.85\linewidth]{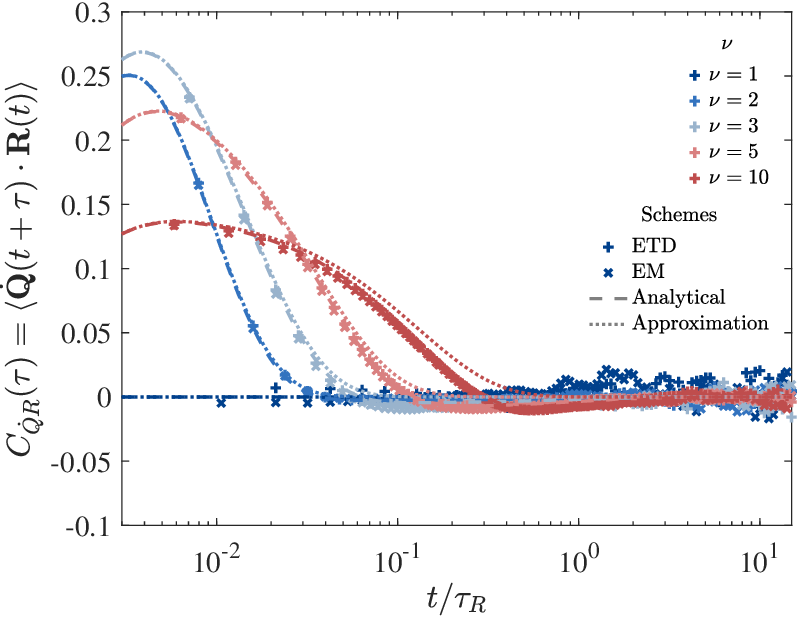}
	\caption{Cross-correlation between the center-of-resistance velocity and the connector vector, $\langle \dot{\mathbf Q}(t+\tau)\!\cdot\!\mathbf R(t)\rangle$, plotted against the scaled time $t/\tau_R$ for several asymmetry ratios $\nu$. Dashed lines show the exact analytical prediction Eq.~\eqref{eq:CQR_final} and dotted lines the short-time approximation Eq.~\eqref{eq:CQR_boundary} (see Appendix~\ref{appendix:CQR}).}
	\label{fig.dumbbellCQR}
\end{figure}
A direct measure of the translational--configurational coupling is provided by the cross-correlation between the drag-center velocity and the connector vector, shown in Fig.~\ref{fig.dumbbellCQR}. 
{This quantity can be obtained analytically from the linear Langevin dynamics (see Appendix~\ref{appendix:CQR}). 
The dynamics form an Ornstein--Uhlenbeck process, so that the two-time correlation can be written as
\begin{equation}
C_{\dot QR}(\tau)=3\,\mathbf e_Q\, e^{A\tau}\,\mathcal C\,\mathbf e_R^T,
\end{equation}
where $A$ is the dynamical matrix and $\mathcal C$ is the stationary covariance matrix.
At short times the spring force can be treated as a perturbation (see Appendix~\ref{appendix:CQR} for the expansion), and the correlation reduces to the closed form
\begin{equation}
C_{\dot QR}(\tau)\simeq
\frac{3k_{\mathrm B}T}{\zeta_1+\zeta_2}
\bigl(e^{-(\zeta_2/m_2)\tau}-e^{-(\zeta_1/m_1)\tau}\bigr),
\label{eq:CQR_boundary}
\end{equation}
showing that a transient correlation arises from the mismatch of inertial relaxation rates of the two beads. It vanishes when the two rates coincide, as happens for equal beads ($\nu=1$), and also in the overdamped limit ($m_i\to0$), where both rates diverge and the translational and configurational degrees of freedom decouple.}
For $\nu>1$, however, a finite positive short-time signal appears, showing that the instantaneous connector orientation biases the subsequent motion of the drag center in the same direction. 
{The analytical solution Eq.~\eqref{eq:CQR_final} agrees closely with the simulation data over the full range of $t/\tau_R$, while the short-time approximation Eq.~\eqref{eq:CQR_boundary} captures the early-time behavior, as expected from its perturbative origin.}
The amplitude of this signal increases with $\nu$, and its decay to zero is shifted to later times, indicating that increasing asymmetry both strengthens and prolongs the inertia-induced coupling between translational and configurational degrees of freedom. 
{This trend follows directly from the structure of the matrix exponential $e^{A\tau}$, whose characteristic timescales depend on the friction contrast between the two beads.}
The ETD and EM results remain in close agreement, confirming that this effect is physical rather than a numerical artifact. 
{The agreement between simulation and analytical prediction further demonstrates that the observed coupling is an intrinsic feature of the underdamped dynamics, rather than a discretization effect.}
Together with the short-time suppression of $C_{R\dot R}(t)$ and the high-frequency modification of the configurational spectrum, this result shows that finite inertia in lopsided dumbbells is expressed most clearly through transient coupling between otherwise distinct translational and internal modes.

\section{Conclusion}\label{sec:conclusion}
We studied finite-inertia effects in Langevin dynamics of a lopsided Hookean dumbbell and assessed exponential-time-differencing (ETD) schemes as stiff integrators in the small-mass regime. 
{In particular, at large timesteps the advantage of ETD is substantial: for example, at $h=0.004$ the Euler--Maruyama scheme incurs an error of about $31\%$ in $\langle R^2\rangle$, whereas ETD2 remains accurate within $0.3\%$.}
Starting from the inertial bead-level equations, we derived the coupled $(\mathbf Q,\mathbf R)$ formulation and used it to obtain analytic benchmarks for both time- and frequency-domain observables.

Across asymmetry ratios $\nu\in[1,10]$, ETD and EM reproduce the correct configurational Boltzmann statistics for the connector length $R=|\mathbf R|$ [Eq.~\eqref{eq.dumbbellPR}], and all schemes agree on the long-time translational diffusivity of the center of resistance, $D_Q=k_BT/(\zeta_1+\zeta_2)$. In the configurational subspace, the time-domain correlators $C_{RR}(t)$ and $C_{R^2R^2}(t)$ collapse onto the overdamped Ornstein--Uhlenbeck predictions when time is rescaled by $\tau_R=\zeta_{\rm eff}/H$, consistent with the strong separation $\tau_m\ll\tau_R$ in the parameter range considered. For the orientational relaxation, we provided a closed-form overdamped reference for the elastic dumbbell [Eq.~\eqref{eq.exact_Cuu}] and found that underdamped simulations collapse onto the same $t/\tau_R$ curve, while inertial constraints only affect the immediate $t\to0$ behavior.
In this sense, the present results clarify the regime of validity of the classical overdamped Langevin description for lopsided dumbbells: it remains accurate for the dominant long-time configurational relaxation, but misses finite-inertia corrections in short-time and high-frequency observables.

Finite-inertia effects at short times are most clearly resolved by diagnostics sensitive to high frequencies and mode coupling. The end-to-end PSD reveals a clear frequency-dependent crossover: the low-frequency regime is governed by configurational relaxation, an intermediate regime remains close to the overdamped scaling, and the high-frequency regime crosses over to the inertial $f^{-4}$ tail implied by the $\omega^{-4}$ asymptotics of the spectrum. Likewise, the position--velocity cross-correlation $C_{R\dot R}(t)$ shows a transient suppression relative to the overdamped prediction when $\nu>1$, with a crossover duration that increases with asymmetry. In addition, the cross-correlation between the drag-center velocity and the connector vector remains nearly zero for the symmetric dumbbell but develops a finite short-time signal for $\nu>1$, demonstrating that finite inertia and bead asymmetry jointly induce a transient coupling between translational and configurational modes. Taken together, these results indicate that finite inertia primarily modifies the shortest-time configurational fluctuations in lopsided dumbbells through high-frequency corrections and transient mode coupling, while the dominant low-frequency relaxation remains close to the overdamped description. 
This framework enables a systematic bridge between overdamped polymer models and finite-inertia mesoscopic descriptions, enabling controlled access to short-time and high-frequency dynamics.

Future work will incorporate hydrodynamic interactions and non-Hookean elasticity, where the $(\mathbf Q,\mathbf R)$ coupling and inertia may influence not only short-time transients but also mesoscale transport and rheological responses.

\begin{acknowledgments}
The authors are also grateful for the support of the Zhejiang Provincial Natural Science Foundation of China (No. LZ25A020005) and the National Natural Science Foundation of China (No. 12572296).
\end{acknowledgments}

\section*{Research ethics}
There is no ethical issue in this submitted manuscript and its authorship.

\section*{Author contributions}
All authors have accepted responsibility for the entire content of this manuscript and approved its submission.

\section*{Conflict of Interest}
The authors have no conflicts to disclose.

\section*{Data Availability Statement}
The data that support the findings of this study are available from the corresponding author upon reasonable request.

\appendix
\section{Analytical expression for $C_{\dot QR}(\tau)$}
\label{appendix:CQR}

We derive the exact analytical expression for the cross-correlation
\begin{equation}
C_{\dot QR}(\tau)=\langle\dot{\mathbf Q}(t+\tau)\cdot\mathbf R(t)\rangle,
\label{eq:CQR_def}
\end{equation}
where $\dot{\mathbf Q}=(\zeta_1\mathbf v_1+\zeta_2\mathbf v_2)/(\zeta_1+\zeta_2)$ is the centre-of-resistance velocity and $\mathbf R=\mathbf r_2-\mathbf r_1$ is the connector vector.

We consider the quiescent case with $\mathbf u_i=0$ in Eq.~\eqref{eq:langevin} and introduce $\mathbf v_i\equiv\dot{\mathbf r}_i$,
\begin{equation}
\begin{aligned}
m_1\dot{\mathbf v}_1 &= -\zeta_1\mathbf v_1 + H(\mathbf r_2-\mathbf r_1) + \mathbf F_1^{(b)}(t),
\\
m_2\dot{\mathbf v}_2 &= -\zeta_2\mathbf v_2 + H(\mathbf r_1-\mathbf r_2) + \mathbf F_2^{(b)}(t).
\label{eq:langevin_app}
\end{aligned}
\end{equation}
Defining the per-dimension state vector $\mathbf Y=(r_1,r_2,v_1,v_2)^T$, the dynamics becomes
\begin{equation}
\frac{dY_\alpha}{dt} = A_{\alpha\beta}Y_\beta + G_\alpha(t),
\label{eq:state_space}
\end{equation}
with summation over $\beta$ implied, and $G_\alpha(t)$ grouping the Brownian forces. The dynamical matrix is
\begin{equation}
A = \begin{pmatrix}
0&0&1&0\\
0&0&0&1\\
-H/m_1&H/m_1&-\zeta_1/m_1&0\\
H/m_2&-H/m_2&0&-\zeta_2/m_2
\end{pmatrix}.
\label{eq:A_matrix}
\end{equation}

Equation~\eqref{eq:state_space} is a linear stochastic differential equation with additive Gaussian noise. Its stationary two-time correlation has the closed-form expression
\begin{equation}
\langle Y_\alpha(t+\tau)Y_\beta(t)\rangle = [e^{A\tau}]_{\alpha\gamma}\,\mathcal C_{\gamma\beta},
\label{eq:ou_corr}
\end{equation}
where $\mathcal C_{\alpha\beta}=\langle Y_\alpha Y_\beta\rangle$ is the stationary covariance. This follows from the standard theory of linear Ornstein--Uhlenbeck processes. By equipartition, $\langle v_i^2\rangle = k_BT/m_i$, $\langle v_1v_2\rangle=0$, and $\langle r_iv_j\rangle=0$. For the spring, the Boltzmann distribution gives $\langle(r_2-r_1)^2\rangle=k_BT/H$, which together with $\langle r_1^2\rangle=\langle r_2^2\rangle$ yields $\langle r_1r_2\rangle=k_BT/(2H)$. The covariance matrix is therefore
\begin{equation}
\mathcal C = \begin{pmatrix}
k_BT/H & k_BT/2H & 0 & 0\\
k_BT/2H & k_BT/H & 0 & 0\\
0 & 0 & k_BT/m_1 & 0\\
0 & 0 & 0 & k_BT/m_2
\end{pmatrix}.
\label{eq:C_matrix}
\end{equation}
The full Lyapunov equation is ill-defined here because of the zero eigenvalue associated with free centre-of-mass diffusion; the physically motivated construction above bypasses this difficulty and yields the correct projection onto the internal subspace.

The relevant quantities are projections: $\mathbf R = \mathbf e_R\mathbf Y$, $\dot{\mathbf Q} = \mathbf e_Q\mathbf Y$, with $\mathbf e_R=(-1,1,0,0)$ and $\mathbf e_Q=(0,0,\zeta_1/(\zeta_1+\zeta_2),\zeta_2/(\zeta_1+\zeta_2))$. Using Eq.~\eqref{eq:ou_corr} and accounting for the three spatial dimensions,
\begin{equation}
C_{\dot QR}(\tau) = 3\,\mathbf e_Q\, e^{A\tau}\,\mathcal C\,\mathbf e_R^T.
\label{eq:CQR_final}
\end{equation}

The companion correlation $C_{R\dot Q}(\tau)=\langle\mathbf R(t+\tau)\cdot\dot{\mathbf Q}(t)\rangle$ follows as $C_{R\dot Q}(\tau)=3\,\mathbf e_R\, e^{A\tau}\,\mathcal C\,\mathbf e_Q^T$ and satisfies $C_{R\dot Q}(\tau)=C_{\dot Q R}(-\tau)$ by time-translation invariance. Equation~\eqref{eq:CQR_final} is evaluated via the matrix exponential for each $\tau$. For $\nu=1$, $\mathbf e_Q\perp\mathbf e_R$ gives $C_{\dot QR}(\tau)=0$ identically.


A transparent closed form is obtained by expanding the dynamics at short times. Write $A = A_0 + H A_1$, where $A_0$ is the dynamical matrix with $H=0$ and $A_1$ contains only the spring-coupling entries. Expanding the matrix exponential as
\begin{equation}
e^{A\tau}=e^{A_0\tau}
+H\int_0^\tau e^{A_0(\tau-s)}A_1\,e^{A_0 s}\,ds
+O(H^2\tau^2),
\end{equation}
and inserting into Eq.~\eqref{eq:CQR_final}, the zeroth-order term vanishes because $\mathbf e_Q\cdot e^{A_0\tau}$ has no projection onto the position subspace spanned by $\mathcal C\,\mathbf e_R^T$. At first order, the factor $H$ multiplying the integral compensates the factor $1/H$ in $\mathcal C$ [Eq.~\eqref{eq:C_matrix}], and evaluating the integral yields
\begin{equation}
C_{\dot QR}(\tau)\simeq
\frac{3k_{\mathrm B}T}{\zeta_1+\zeta_2}
\bigl(e^{-(\zeta_2/m_2)\tau}-e^{-(\zeta_1/m_1)\tau}\bigr),
\label{eq:CQR_free_particle}
\end{equation}
which isolates the inertial relaxation of each bead. For $\nu=1$ the two rates coincide and Eq.~\eqref{eq:CQR_free_particle} vanishes identically.

\bibliography{references.bib}

\end{document}